\documentclass[graybox]{svmult}

\usepackage{type1cm}        
\usepackage{makeidx}         
\usepackage{graphicx}        
\usepackage{multicol}        
\usepackage[bottom]{footmisc}

\usepackage{newtxtext}       %
\usepackage[varvw]{newtxmath}       

\makeindex             

\begin{document}

\title*{TrueReason: An Exemplar Personalised Learning System Integrating Reasoning with Foundational Models}
\titlerunning{TrueReason: Integrating Reasoning with Foundational Models} 
\author{Sahan Bulathwela\orcidID{0000-0002-5878-2143} , Daniel Van Niekerk\orcidID{0000-0002-7324-2751}, 
\\Jarrod Shipton\orcidID{0000-0003-3590-9671} ,  Maria Perez-Ortiz\orcidID{0000-0003-1302-6093},
\\ Benjamin Rosman\orcidID{0000-0002-0284-4114} and John Shawe-Taylor\orcidID{0000-0002-2030-0073}}
\authorrunning{Bulathwela et al.} 

\institute{Sahan Bulathwela, Daniel Van Niekerk, Maria Perez-Ortiz and John Shawe-Taylor  \at Centre for Artificial Intelligence, Department of Computer Science, University College London \email{\{m.bulathwela, maria.perez, j.shawe-taylor\}@ucl.ac.uk}
\and Jarrod Shipton and Benjamin Rosman \at School of Computer Science and Applied Mathematics, University of the Witwatersrand \\ \email{\{jarrod.shipton, benjamin.rosman1\}@wits.ac.za}
}

%
\maketitle

\abstract*{Personalised education is one of the domains that can greatly benefit from the most recent advances in Artificial Intelligence (AI) and Large Language Models (LLM). However, it is also one of the most challenging applications due to the cognitive complexity teaching effectively while personalising the learning experience to suite independent learners. We hypothesise that one promising approach to excelling such demanding use-cases is using a \emph{society of minds}. In this chapter, we present TrueReason, an exemplar personalised learning system that integrates a multitude of specialised AI models that can mimic \emph{micro skills} that are composed together by a LLM to operationalise planning and reasoning. The architecture of the initial prototype is presented while describing two micro skills that have been incorporated in the prototype. The proposed system demonstrates the first step in building sophisticated AI systems that can take up very complex cognitive tasks that are demanded by domains such as education.}

\abstract{Personalised education is one of the domains that can greatly benefit from the most recent advances in Artificial Intelligence (AI) and Large Language Models (LLM). However, it is also one of the most challenging applications due to the cognitive complexity teaching effectively while personalising the learning experience to suite independent learners. We hypothesise that one promising approach to excelling such demanding use-cases is using a \emph{society of minds}. In this chapter, we present TrueReason, an exemplar personalised learning system that integrates a multitude of specialised AI models that can mimic \emph{micro skills} that are composed together by a LLM to operationalise planning and reasoning. The architecture of the initial prototype is presented while describing two micro skills that have been incorporated in the prototype. The proposed system demonstrates the first step in building sophisticated AI systems that can take up very complex cognitive tasks that are demanded by domains such as education.}



\section{Introduction}
\label{sec:intro}

The use of Artificial Intelligence (AI) in education has caught the attention of many computer scientists and educators in the last few years. In AI-supported education, sub-topics range from personalised learning to Feedback Generation to Retention and Learner Success modelling. In personalised learning systems, Educational Recommendation Systems (EdRecSys) cover a wide variety of systems that support informal learners to acquire knowledge from suitable learning resources. With the rapid increase of video lectures and Open Educational Resources \cite{bulathwela_edm_population}, the abundance of learning resources available in the Internet increases the value of EdRecSys which can use these materials to create personalised learning pathways.   

Access to education benefits individuals and societies alike and goes hand-in-hand with social inclusion, economic growth, poverty reduction and equality. AI-enabled learning applications getting better over time helps in scaling education, leading to the mass democratisation of learning opportunities  \cite{su16020781}. The upward trend of AI in Education stems greatly from the incorporation of generative AI technologies and foundational AI models such as Large Language Models (LLMs) in AI-enabled educational applications \cite{denny2024generative}. Despite their impressive performance on a wide variety of tasks, LLMs fall short on tasks that involve complex reasoning, argumentation and planning. This limitation significantly hinders the expressiveness we require from a human-like learning companion that would be required to enrich learning journeys.  

One of the most ambitious use cases of computer-assisted learning is personal recommender systems for lifelong learning which require sophisticated recommendation models accounting for a wide range of factors such as background knowledge of learners and novelty of the material while effectively maintaining knowledge states of learners for significant periods \cite{truelearn}. A foundational component in such systems is a model that captures and tracks learners’ knowledge states to assist them on their path of knowledge discovery and acquisition. One such model – TrueLearn – uses Bayesian algorithms and defines knowledge components in terms of Wikipedia topics to provide transparent estimates of learners’ knowledge states \cite{truelearn,truelear_python}. While models such as TrueLearn could form the backbone of a lifelong learning recommendation system by tracking learners’ progress through engagement with open educational resources, many aspects and challenges associated with such a personal recommender are left unaddressed – not least the model “cold-start problem” which is relevant when onboarding a new learner \cite{bulathwela_edm_population}.

\subsection{Chapter Overview}

This chapter describes the TrueReason system. Section \ref{sec:edrecsys} informs the prior work setting the context to the proposal by describing the different components of a personalised learning system and linking it to the current advances in LLMs that motivate the design choices utilised. Section \ref{sec:truereason} describes the proposed exemplar personalised learning system with its subcomponents. This is followed by section \ref{sec:microskills} where two of the integrated micro skills, namely reinforcement learning-based multi-step recommender (section \ref{sec:Reinforcement_Learning_ER}) and topic-controlled question generation (section \ref{sec:qg}) are described in detail with specific results. The broader discussion that links the proposed method and its ability to create sophisticated learning companions while breaking sustainability and accessibility barriers to the system is done in section \ref{sec:dis} and the chapter is concluded in section \ref{sec:conclusion}.

\section{Personalised Learning Systems} 
\label{sec:edrecsys}

A wealth of research has been conducted on improving personalised learning systems. Among the applications that utilise personalised learning, intelligent tutoring systems and educational recommenders dominate the literature. In this section, we review and discuss in detail the different components of a personalised learning system. 

\subsection{Components of a Personalised Learning System}
\begin{figure}[]
    \centering
    \includegraphics[width=\textwidth]{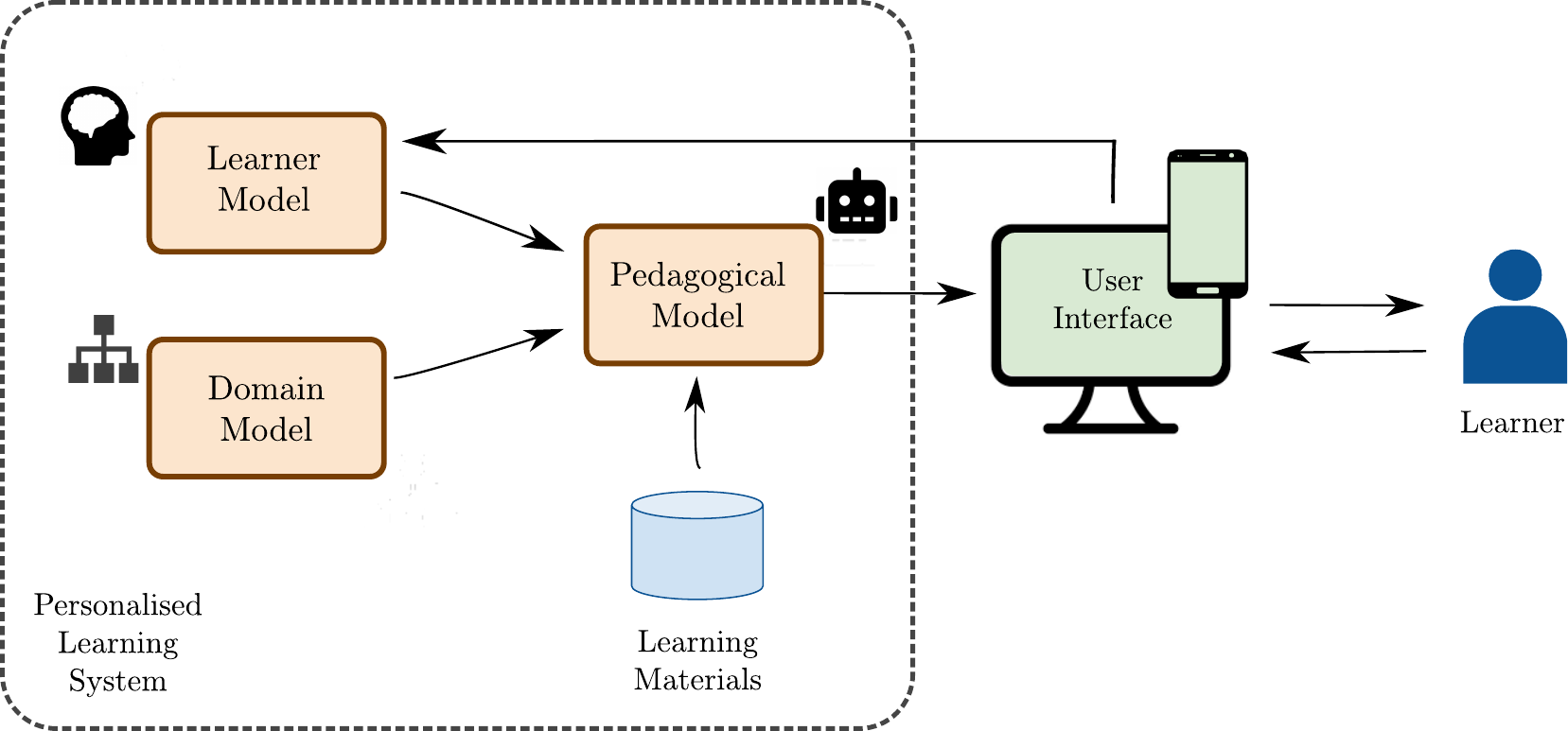}
    \caption{
    The architecture of a personalised learning system that interacts with the learner using a user interface. 
    }
    \label{fig:its_figure}
    \end{figure}

 A personalised learning system incorporates different components that interact with each other to recommend suitable learning trajectories to learners. These core components of a conventional personalised learning system can be seen in Figure \ref{fig:its_figure}. 

\subsection{Domain Model}
For any knowledge area, it is possible to define a scope of specific topics and subtopics that are relevant when learning. Such a structure also allows for identifying prerequisites that need to be covered earlier in the lesson sequence to provide the foundational knowledge complementary to understanding more advanced topics. Such a structure that can be exploited in creating effective lesson plans is identified as a domain model. 

In an educational system, \textbf{Knowledge Components (KCs)} are atomic units of learnable knowledge that can be learned and mastered. These KCs can come from a standard taxonomy or generated specifically for an application  \cite{Corbett1994}. The primary approach used for KC tagging is expert labelling. This method is very accurate and aligns well with standards and frameworks set out by stakeholders. However, expert labelling is resource-intensive and limiting in terms of scalability. In a more relaxed setting where user interest is captured, social tagging is another approach that has been used \cite{klavsnja2018social,sampson2011ask} which can be prone to errors due to the contributions from non-experts. Automating content annotation has been proposed, although methods usually rely on partial expert labelling \cite{Lindsey2014,nn_know_comp}, which is still resource intense, or use unsupervised topic modelling (e.g. Latent Dirichlet Allocation \cite{blei2003latent}) which entails complex hyperparameter tuning with no guarantee of identifying humanly intuitive topics. 

Entity Linking \cite{Brank2017,wat_api} is a more recent approach that can provide both scalability and humanly intuitive representations. In a nutshell, this approach links parts of free-form text to entities in an existing knowledge base (such as Wikipedia). Prior work has proposed informational and educational recommendation systems successfully \cite{piao_social_networks,piao_mooc}. Entity linking poses the risk of using entities that may not necessarily align with an education expert's handcrafted taxonomy. In the proposed system, we heavily rely on entity linking (specifically, Wikificaiton \cite{Brank2017}) as the system needs to 1) work with a large, evolving collection of learning materials (requiring scale), and 2) a wide array of topics (to cover diverse topics in lifelong learning scenarios).

In contrast to predicting the personalised \emph{next} learning material (e.g. \cite{truelearn,piaopredicting}, planning longer-term learning journeys that entail a multitude of topic transitions is more complex and is often underexplored. Again, solutions resort to expert annotation, where experts manually define sensible topic transitions in a handcrafted taxonomy. While this method can work for personalisation within very specialised knowledge areas with a narrow scope \cite{bauman2018recommending}, defining relationships between topics at scale is essential for building a lifelong learning system. Due to this reason, using universal knowledge bases such as Wikipedia, that evolve while supporting multilingualism has great potential \cite{su16020781}. Apart from the topic coverage, another advantage of using a knowledge base like Wikipedia is that there is auxiliary information about semantic relatedness \cite{ponza_semantic_relate} between topics and hierarchical structures of topics (ie. topic $a1$ and $a2$ are subtopics of topic $A$). Our prior work also has heavily relied on using Wikipedia as a domain model for universal knowledge \cite{truelearn,su16020781} while some works have specifically explored how properties like semantic relatedness can be exploited to improve personalisation \cite{semantic_truelearn}. Our prior works have also shown how Wikipedia-based topics can lead to humanly intuitive explanations of learner state \cite{truelear_python} and content representations \cite{x5learn}. Due to these advantages, we also make use of Wikipedia as a source for the domain model within this work.        

\subsection{Learning Resources} 

While the domain model defines the structure of knowledge, learning materials/resources are required to carry out the actual learning. Therefore, a core component of any educational system is the collection of educational resources. Learning materials can come in different modalities (e.g. video/audio/text etc.),
and different languages entailing many types of activities (e.g. questions, instructional videos, lesson summaries, interactive games etc.).
The early definition of interactive educational resources was limited to dynamic modalities such as simulations and educational games. With the advent of AI, the scope of this definition has expanded significantly where AI-enabled innovations such as intelligent textbooks have also joined. The diversity of learning materials is important to provide novel learning experiences while increasing learner engagement. 

While in-person instructional lecturing has been a main form of knowledge transfer from the inception of education, adapting this teaching method to the digital realm has led to educational videos. With the mass availability of Internet connectivity and the popularisation of platforms such as YouTube, we find 100,000s of educational videos being created and being made available via the Internet. The mass availability of video-based educational resources was further expanded with the introduction of Massive Open Online Courses (MOOCs). Due to their heavy presence and effect, educational videos are extensively researched in the literature. A wealth of research attempts to build educational recommendation systems around video collections. Many work studies learner engagement with videos while others propose good practices to create engaging educational videos. Our prior work also has studied building educational recommendation systems for educational videos including releasing public datasets based on learner engagement with educational videos such as PEEKC and VLE datasets. This work builds the final recommender using a collection of educational videos from the PEEKC dataset~\cite{bulathwela2021peek}.         

\subsection{Learner Model} 

A learner model is a representation of the learner's state. This can be based on a multitude of verticals. A few frequently used verticals are learner knowledge/skill mastery \cite{Corbett1994,Yudelson13,SyedC17} and interests/goals \cite{zarrinkalam2020extracting,bulathwela2022sus}. We can hypothesise the relationship between learner engagement and the learner state using the probability distributions presented via equations \ref{eq:know} and \ref{eq:int} where learner $\ell$ interacts with learning materials $r_x$ that consists of knowledge components $K_{r_x}$. The latent knowledge and interest states of the learner $\ell$ at time $t$ are $\theta^t_{\ell_{\texttt{K}}}$ and $\theta^t_{\ell_{\texttt{I}}}$ respectively.

\begin{equation} \label{eq:know}
 P(\theta^t_{\ell_{\texttt{K}}} | e^{t}_{\ell,r_x}, K_{r_x}) \propto P( e^{t}_{\ell,r_x} | \theta^t_{\ell_{\texttt{K}}}, K_{r_x}) \cdot P(\theta^t_{\ell_{\texttt{K}}})
\end{equation}

\begin{equation}  \label{eq:int}
 P(\theta^t_{\ell_{\texttt{I}}} | e^{t}_{\ell,r_x}, K_{r_x}) \propto P(e^{t}_{\ell,r_x} | \theta^t_{\ell_{\texttt{I}}}, K_{r_x}) \cdot P(\theta^t_{\ell_{\texttt{I}}})
\end{equation} 

Modelling the knowledge and interest state of the learners can be done in both sub-symbolic and neural approaches. The utilisation of deep neural networks in modelling learner knowledge state (a.k.a. \emph{ deep knowledge tracing}) has led to significant breakthroughs in recent years evolving from the use of recurrent neural networks \cite{deep_kt} to transformers \cite{shin2021saint+}. Neural representations derive a distributed representation of the learner that is not as humanly intuitive. However, due to the neural networks' ability to carry out end-to-end learning, there is flexibility to learn very effective features \cite{xiong2016going}. We direct the reader to \cite{abdelrahman2023knowledge} to get an extensive understanding of the current trends in knowledge tracing. Similar trends are evident in interest modelling where deep learning has pioneered the recent advances in the education domain \cite{piaopredicting,piao_mooc}. While neural models demonstrate superior predictive performance, they pose challenges when interpreting/explaining their predictions. This can become a setback when it comes to creating human-centric AI systems where we intend to seamlessly integrate AI systems with humans as the challenges in explainability introduce communication barriers between the AI system and the consumer. At the same time, they require a large amount of data to train them also adding to computational costs. The state-of-the-art models are also huge, introducing unique cost and complexity challenges when maintaining model infrastructure.

In contrast to neural models, \emph{symbolic representation} of learner state entails using a taxonomy that is familiar to human understanding (ie. a formal symbol system). Techniques such as Bag-of-Concepts and Probabilistic Graphical Modelling use a sub-symbolic model approach where the representation is humanly-intuitive. Concept-based user modelling (a.k.a. Bag-of-Concepts) \cite{zarrinkalam2020extracting} is a commonly used technique in this domain where expert annotation \cite{Corbett1994} or automatic concept extraction using methods like topic classification \cite{KANG201752} and entity linking \cite{bulathwela2022sus} are used to tag concepts to content. A more advanced modelling approach is to build a Bayesian graphical model to present the data generation process within knowledge acquisition (a.k.a. Bayesian Knowledge Tracing \cite{Corbett1994}). The variety of models that provide humanly intuitive representations of human learners is called \emph{open learner models (OLM)} \cite{Bull2016}. Many works are showing that the familiarity brought by OLMs can help learners with metacognition \cite{bull2008metacognition} and self-regulated learning \cite{hooshyar2020open}. Having symbolic representations heavily complements their ability to collaborate with human learners as the representation can be presented to the learner in the form of a visualisation that is easily understood. 

While there has been extensive work done on modelling human learning on both Bayesian and neural fronts, many focus on the test-taking scenario where knowledge state inference is done based on how learners answer questions \cite{Yudelson13,deep_kt}. While such explicit feedback is more accurate, it is hard to obtain in an informal lifelong learning system as question answering hinders user experience. Hence, informal learning systems should rely on implicit feedback such as dwell time and clicking to infer knowledge acquisition and interest. Recent learner models built for such systems in Search as Learning (SAL) \cite{SyedC17} and lifelong learning applications \cite{truelearn} rely on knowledge tracing-inspired models that rely on implicit feedback.  

\emph{TrueLearn} is a state-of-the-art probabilistic graphical model that can model learner interest and knowledge state using implicit engagement signals \cite{bulathwela2022sus}. Its humanly intuitive learner representation and data efficiency and availability in the form of a Python package make it highly applicable to building lifelong learning platforms \cite{truelear_python}. Inferring these latent variables explicitly informs the planning of lesson personalisation. Due to these reasons, we use TrueLearn as the learner model in our work.

\subsection{Pedagogical Model} 


Even with the presence of key components such as a learner model, a domain model, and a comprehensive database of learning materials, a well-designed teaching strategy remains crucial for sequencing the most appropriate learning materials for each individual learner. This strategy serves as the guiding framework to ensure that educational content is not only relevant but also delivered in a manner that maximises the learner's engagement and understanding.

A teaching strategy can draw from various pedagogical models and adapt to different learning preferences. For example, approaches like Vygotsky's zone of proximal development can help guide learners through tasks that are just beyond their current ability but can be mastered with appropriate support. Previous personalised models have e.g. integrated pedagogical theories of development \cite{truelearn}. Similarly, strategies can account for different cognitive or sensory learning styles, such as visual, auditory, kinesthetic, or reading/writing preferences, tailoring the instructional methods to better align with how each learner absorbs information.

These preferences and pedagogical models can also be refined dynamically, based on the learner's ongoing interactions with the system. For instance, as learners engage with the platform, it can track their progress, analyse their responses, and adjust the instructional sequence or methods to optimise their learning journey. This adaptability ensures that the teaching strategy evolves in real-time, providing increasingly personalised learning pathways.

Moreover, learners often have distinct preferences regarding the modalities through which they consume educational content. Some may prefer watching instructional videos, others might benefit from reading detailed texts or engaging with interactive simulations. By integrating a variety of content formats—such as videos, infographics, text, quizzes, or hands-on activities—the teaching strategy can cater to diverse learner needs and ensure a richer, more inclusive learning experience.

Incorporating these factors—pedagogical models, learning preferences, and content modalities—not only makes the learning process more personalised but could also enhance learner motivation and retention. A well-constructed teaching strategy has the potential to make education more inclusive by recognising the unique strengths, challenges, and preferences of each learner, ensuring that no one is left behind and that all learners are supported in achieving their full potential.

\subsection{User Interface} 


The final component, the user interface, is essential for bridging the intelligence of the personalization system with the human learner. It serves as the primary point of interaction, allowing users to engage with a highly complex personalized learning environment in a simple and intuitive manner. While the system itself may consist of numerous sub-components—such as recommendation engines, content delivery systems, and real-time data analysers—the user interface is responsible for presenting this complexity in a way that feels seamless and accessible to the learner. By obfuscating the underlying technological intricacies, it allows learners to focus solely on their educational goals without being overwhelmed by the system’s internal workings.

An effective user interface is not just a functional necessity; it is an integral part of the overall learning experience. It determines how learners navigate through materials, how they receive feedback, and how they interact with adaptive features that cater to their unique needs. A well-designed interface can significantly enhance engagement, motivation, and learning outcomes by making the system feel responsive, personalized, and user-friendly.

Innovative approaches have emerged to refine this critical aspect of personalized learning systems. Designers and developers now prioritize creating interfaces that are not only functional but also intuitive, adaptive, and visually appealing. Key strategies include minimizing cognitive load by reducing unnecessary complexity, providing clear navigation and guidance, and offering real-time support or hints as learners progress through tasks. The goal is to make the system feel as natural and effortless to use as possible, encouraging learners to explore and engage deeply with the content.

One particularly promising recent innovation in this area is the use of large language models (LLMs) to enable conversational interaction and intelligent search. LLMs allow for a more human-like interaction between the learner and the system. Rather than navigating through rigid menus or predefined pathways, learners can simply ask questions, request specific content, or seek clarification using natural language. 

In summary, the user interface is far more than a superficial design layer; it is the conduit through which learners experience the full capabilities of a personalized learning system. By making the interface seamless and incorporating cutting-edge technologies like large language models, developers can create an environment that not only adapts to individual learning needs but also feels intuitive and engaging, empowering learners to achieve their educational objectives with greater ease.

\subsection{Large Language Models and Knowledge Intense Tasks}

With the recent advancement of deep learning, the SOTA of many applications fronts have shifted to deep neural models. Among these advances, LLMs have made pivotal movement in revolutionising knowledge intensive tasks such as education. Many tasks in education relating to intelligent tutoring \cite{piro2024mylearningtalk,park2024empowering}, personalised recommendation \cite{li2023evaluating} as well as generative tasks such as question generation \cite{bulathwela2023scalable}, feedback generation \cite{stamper2024enhancing,jia2024llm} and summarisation \cite{cachola-etal-2020-tldr} has had breakthroughs. However, these generation models need to excel contextualising/personalising generations to the student state (e.g. weak topics, vocabulary complexity etc.) in order to be effective within a personaliser. The topic-controlled question generation micro-skill described in section \ref{sec:qg} presents a novel approach to teach generic question generators to contextualise generations to topics.

Regardless or the strength of SOTA LLMs, building a holistic personalisation system that can incorporate a multitude of learning activities and pedagogical interventions remains an open question due to the variety of skills such a system demands. Additionally, LLM-powered interfaces can assist in search functions, enabling learners to find relevant resources quickly and accurately by understanding the intent behind their queries. For example, a learner could ask, "Can you explain photosynthesis in simple terms?" or "Show me a video on how to solve quadratic equations," and the system can intelligently interpret and respond to these requests. This eliminates the need for precise keyword searches and enhances the overall accessibility of the system, particularly for users who may not be tech-savvy. 

When pushing the frontiers of conversational information retrieval tasks (including search and question answering), one recent approach that has gained interest is Retrieval Augmented Generation (RAG), where an external document collection is exploited by the LLM to supervise generation \cite{miladi2024comparative,modran2024llm,thus2024exploring}. More abstract interpretations of RAG, such as Toolformer \cite{schick2024toolformer} extends the idea of external help from support documents to tools and APIs. Very recent work in conversational search has also started incorporating the idea of having a collection tools that leads to a more enriched conversation \cite{shi2024learning,shi2024chain}. The overarching approach taken in creating TrueReason (described in section \ref{sec:truereason}) utilises the \emph{society of minds} concept that is inspired by the power of having a swarm of diverse AI/non-AI subcomponents having the ability to compose together to operationalise complex cognitive behaviors.

\section{TrueReason System} \label{sec:truereason}

Many educational recommendation systems and personalised learning platforms today operate within a low bandwidth of learning activities. This means the variety of skills an educational recommender uses is limited to a single activity like recommending videos, web pages or exercises and seldom utilises a multitude of skills such as recommending a video $\rightarrow$ presenting a quiz $\rightarrow$ giving feedback. 
However, a good human teacher would break beyond such a narrow band and incorporate a diverse set of learning activities in variable combinations to make a rich and engaging learning experience. Where self-regulated learning is critical (e.g. informal learning), contextualising teaching strategy can lead to success. \emph{TrueReason Assistant} is a personalised learning system that attempts to build a revolutionary learning companion. It is one of the first works that attempts to fuse foundational models with reasoning (using other specialised AI models). 

\subsection{Problem Background}

TrueReason serves as demonstration of how key limitations found in many educational recommenders can be addressed within the same system. Some of the issues TrueReason addresses are:


\begin{enumerate}
    \item \textbf{Learner On-boarding:} When a new learner begins to use the personal recommender, a new model must be initialised to represent the learner’s knowledge state as accurately as possible to provide relevant recommendations early on. In this case it is useful to elicit learners’ background knowledge in a more direct way than observing engagement and use an alternative recommendation strategy until TrueLearn has a robust estimate from engagement events~\cite{bulathwela2021truelearn}.
    \item \textbf{Learning Goals:} To assist the learner beyond recommending novel and engaging educational content, it is useful to base interactions on a basic situation model which firstly keeps track of learning goals but also allows the learner to monitor and steer their own progress by providing relevant information and feedback about knowledge components and their own knowledge state. This augments learners’ agency and allows them to take control of their learning trajectory instead of being passive receivers of recommendations~\cite{reicherts2022}.
    \item \textbf{Content Ontology:} In contrast to traditional courses for more short-lived educational scenarios, lifelong learning supported by open educational resources has to contend with a more diverse and rapidly changing set of resources which are likely not designed to compose as do lectures in a more formal setting. However, being able to inform learners according to the structure and relations between materials could assist them on their learning path~\cite{ilkou2021}.
    \item \textbf{Knowledge Review:} While using implicit feedback (dwell time and consumption behaviour) is data efficient, having explicit self-testing opportunities helps 1) inculcate self-awareness within the learner and 2) help the learner model to verify the precision of its knowledge state estimates. Appropriate question generation models~\cite{bulathwela2023scalable} could be used to allow learners to test their knowledge or serve as a source of information to initialise or verify the learner knowledge state estimated by TrueLearn. In this direction, personalised question generation can help the AI assistant to create ad-hoc tests to individuals. 
    \item \textbf{Knowledge Gaps:} Because the available set of educational resources are not necessarily designed to be comprehensive combined with the absence of a formal educator, learners may find that knowledge gaps persist after engaging with materials or that they need clarity on unfamiliar concepts. These gaps or explanations could be addressed by world knowledge contained in a large language model or methods for generating analogical explanations~\cite{10.1162/tacl_a_00688} which rely on learners’ background knowledge.

\end{enumerate}

\subsection{System Architecture}
\begin{figure}[b]
        \centering
    
        \includegraphics[width=.7\textwidth]{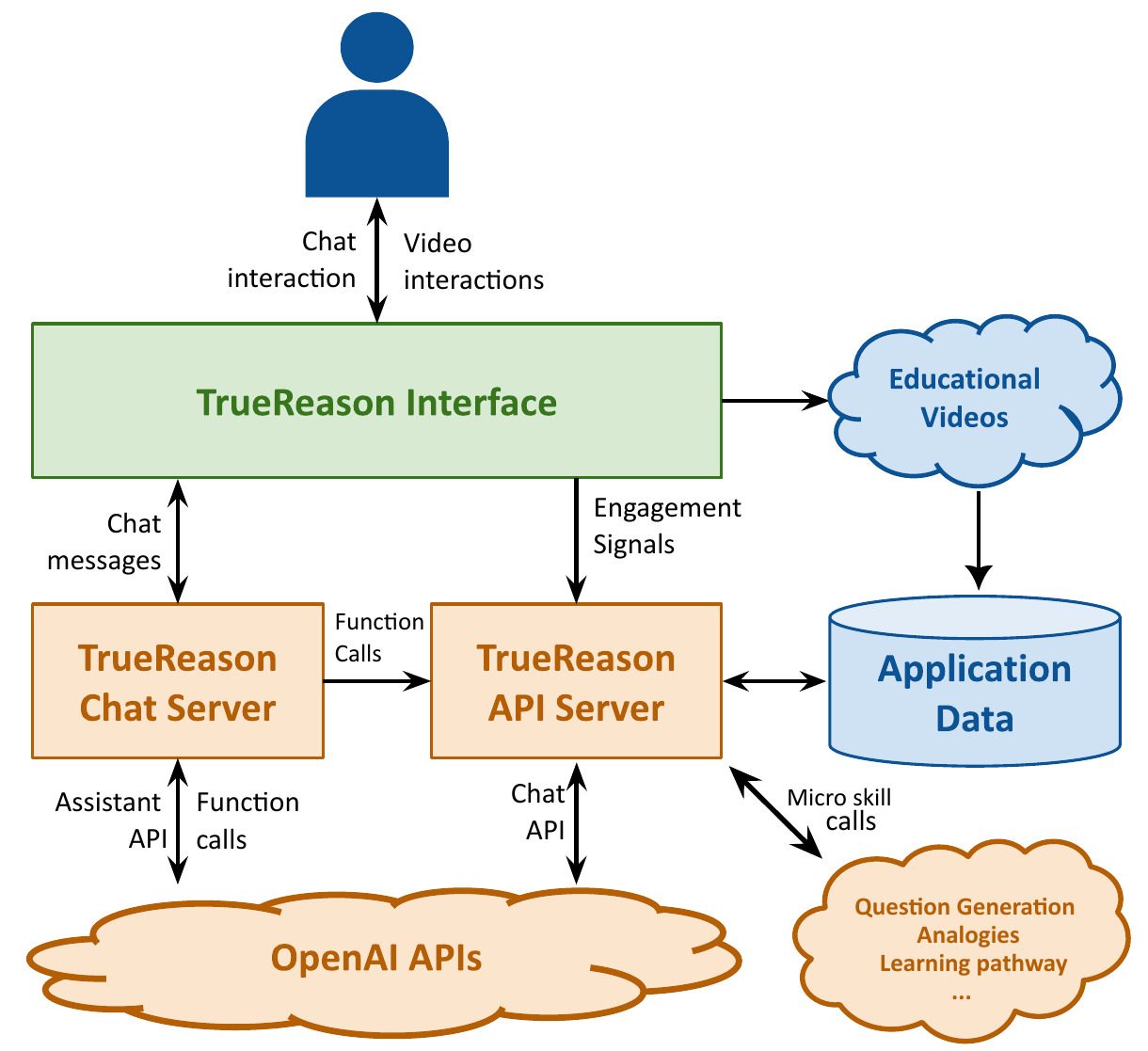}
        \caption{
        The technical architecture of the TrueReason system coordinates the interaction between an intelligent assistant that orchestrates the core conversation of the learner and the API coordinator that acts as the interface between the TrueReason core and the micro-skills. 
        }
        \label{fig:TR_architecture}
    \end{figure}

The overall architecture of the TrueReason system is presented in Figure \ref{fig:TR_architecture}. The system uses a chat interface to interact with the learner. Behind the user interface, the chat component (Chat server) handles the conversation with the learner while it has access to different \emph{skills} (micro-skills to be more precise) that it can utilise to make the conversation more systematic, structured and dynamic. Access to micro-skills (such as query learner state, generate personalised questions, create a learning pathway etc.) is provided through the \emph{TrueReason API Server}.  In this iteration, the TrueLearn model \cite{truelear_python} is used and 
we use a set of educational videos as a collection of learning resources.  

\begin{figure}[]
        \centering
    
        \includegraphics[width=\textwidth]{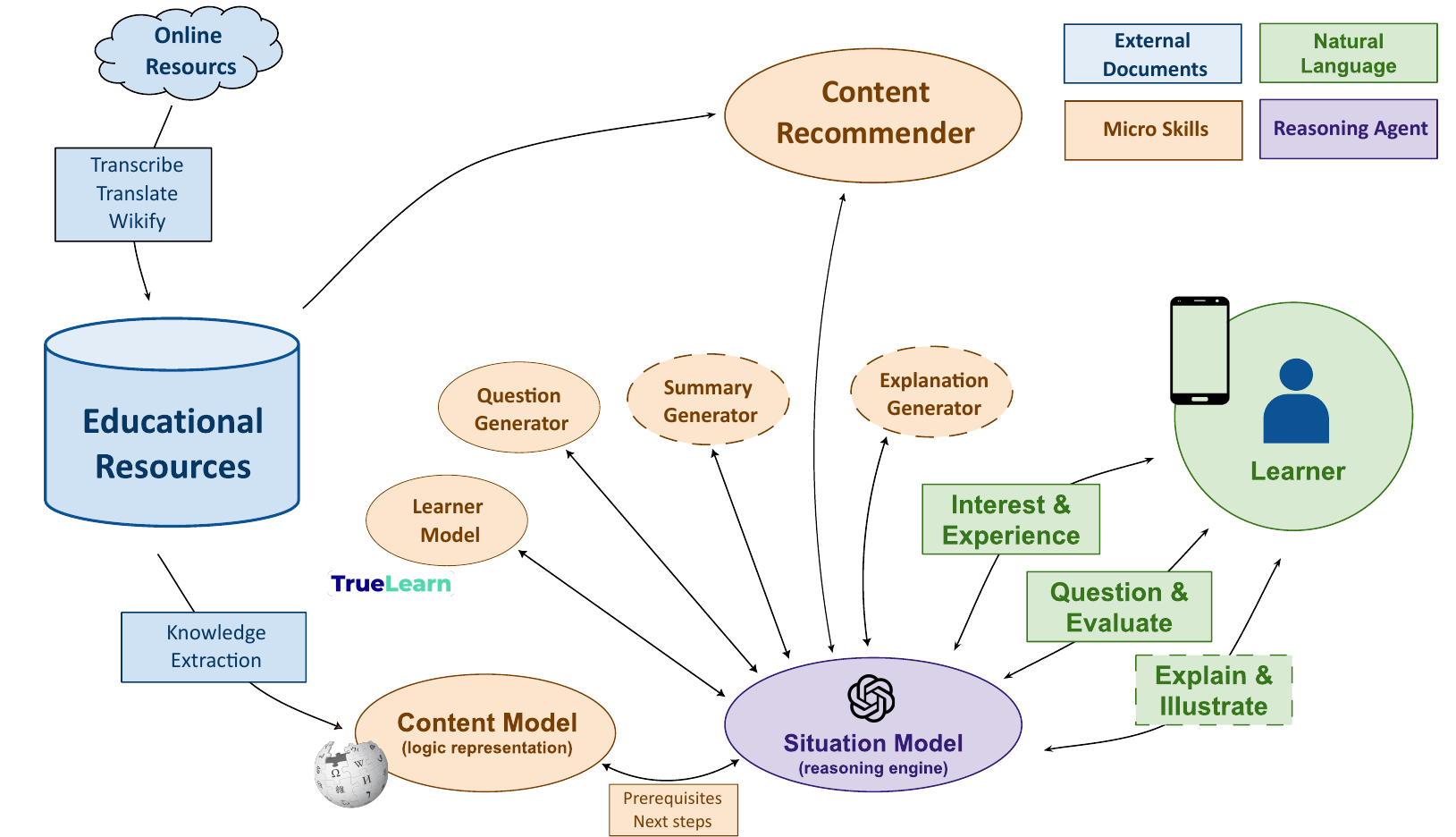}
        \caption{
        The logical components of the TrueReason learning assistant consist of the blue parts that are learning resources, the orange components belong to the core AI assistant that entails 1) the domain model (that is mapped to Wikipedia topics), 2) the learner model, and 3) the pedagogical model, which includes the engagement model and the situation model that interacts with the learner and the green part represents the user interface. The components (e.g. micro skills) that are not implemented at present use dashed lines. 
        }
        \label{fig:logical_model}
    \end{figure}

Contrary to conventional AI-enabled educational systems, TrueReason can incorporate a diverse set of activities within the learning pathway. Figure \ref{fig:logical_model} outlines the logical structure of the TrueReason assistant. As per the figure, the system uses different forms of two-way interaction such as recommendation, retrieval, extracting interest and experience, questioning and evaluation etc. leading to a rich interaction between the learner and the system.

\subsection{Domain Model} \label{sec:domain_model}

As is the practice in TrueLearn, the content model relies on Wikification to extract knowledge components (KCs) from online educational resources and represent them as sets of Wikipedia topics \cite{Brank2017}. Each entry or page in Wikipedia could be considered a valid KC and any resource can be described as a sequence of such sets of topics along with a measure of overlap between the coverage in the resource and the entry in Wikipedia. Segmentation of the resource can be done along defined sections in a document or by taking fixed length slices of videos and audio – see the PEEKC Dataset \cite{bulathwela2021peek} for an example and a description of the conventions also followed here. To construct the content model given an arbitrary collection of educational materials we segment each resource individually and apply Wikification as described above.

\subsubsection{Modelling Relatedness Between KCs}

\begin{figure}[t]
    \centering

    \includegraphics[width=\textwidth]{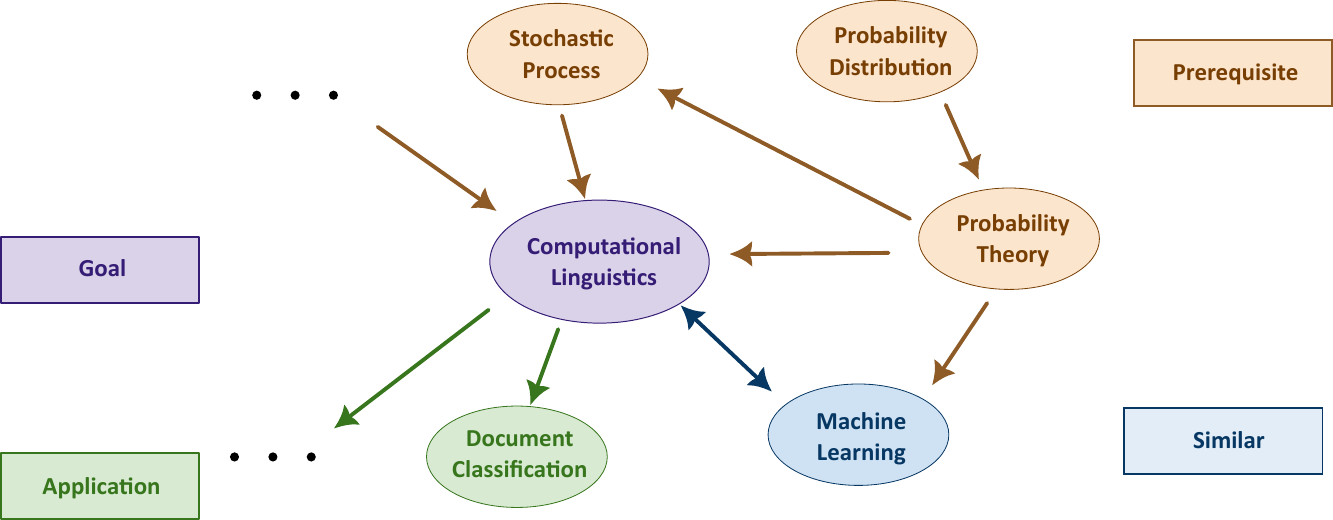}
    \caption{
    Given a goal topic the KC graph over all the content is pruned to a smaller number of related topics which is subsequently labelled by GPT-4 according to the set of relations (shown in red, green, and blue) including prerequisites to and applications of the focal topic (shown in navy blue).}
    \label{fig:domain_model}
\end{figure}

 A weighted directed graph for each resource is then constructed where a link is created between each KC and all those following it in subsequent segments in the same learning resource. The weight of the link is determined by the frequency of one KC being followed by another.  All individual resource graphs are then merged by aggregating the link weights to form a single KC graph which contains information about all the topics in the collection – both the most common sequences reflected by the relative link weights and centrality reflected by the number of outgoing links. Finally, we would like to use this information to retrieve a relevant set of related topics given a topic of interest. We do this on the fly by pruning the KC graph around the topic of interest using their hierarchical level and influence centrality coefficients \cite{moutsinas2021graph} and then applying labels to the links from the set of relations \texttt{is\_prerequisite, is\_application, is\_similar, is\_field} using GPT-4\footnote{\url{https://platform.openai.com/docs/models/gpt-4-turbo-and-gpt-4}}. Figure \ref{fig:domain_model} shows an example of such a pruned graph and relations to the focal topic.

The demonstration content prepared for the initial version of TrueReason Assistant consists of a selection of 1200 videos in the domain of natural language processing hosted by VideoLectures.net\footnote{\url{https://videolectures.net/ }} and the KC graph was constructed based on the PEEKC Dataset which has the same source \cite{bulathwela2021peek}. 

\subsection{Situation Model}

The situation model is responsible for all interactions with new and ongoing learners including dialogue, recommendation, engagement, and knowledge review. To implement these four broad aspects, the state about each need has to be tracked continuously. As per figure \ref{fig:TR_architecture}, this is implemented by a large language model (LLM) assistant (ChatGPT at present) combined with:
\begin{enumerate}
    \item Tools external to the core system that implement specific functionality (\emph{micro skills}) such as multi-step recommendation or question generation – e.g. see section \ref{sec:qg} about contextualised education question generation.
    \item Persistent state capturing learners’ goals, past engagement, and previously recommended resources.
    \item Content index with links to video fragments, stored Wikpedia concept-based representations with text embeddings, and raw transcripts.
\end{enumerate}


Each of the core aspects tracked by the system is described in the following sections.

\subsubsection{Dialogue States}

Dialogue with the learner can be thought of as cycles between three 3 states broadly associated with (1) clarification of learner background and goals, (2) engagement with a recommended resource, and (3) knowledge review and explanation. 
However, these states are only implicit in the instructions provided to the LLM assistant and so learners’ are afforded considerable latitude in their interactions and requests – the state is essentially captured by the dialogue message history.

\subsubsection{Revision of Learner Background and Goals}

Initial interaction is aimed at eliciting learners’ background knowledge and topics of interest in a two-way dialogue which makes use of the content model to inform the learner of related topics and prerequisites they may need to be aware of. This is facilitated by the LLM assistant prompting the learner and extracting (Wikipedia) topics of relevance from their responses to call functions that add, remove, or retrieve KCs that represent learner interest and goals in the application database. In order to present related topics, the assistant has to call a function which queries the domain model as described in Section \ref{sec:domain_model} and translates the result into natural language. 

\begin{svgraybox}

Taking the example in Figure 3, the dialogue may proceed as follows:\\

\textbf{Learner:} I’m interested in knowing more about computational linguistics\\

\textbf{Assistant} calls \texttt{get\_related\_topics}(Computational Linguistics) \\

\textbf{Assistant:} Understanding Computational Linguistics depends on having a grasp of Probability Theory and Stochastic Processes and is often applied in Document Classification. Do you need to review any of the fundamentals or are you more interested in focussing on practical applications?\\

This dialogue typically continues until the user asks for a video recommendation or the assistant suggests the same and the learner agrees.
\end{svgraybox}
    
\subsubsection{Knowledge Review and Explanation}
Once a learner has engaged with a resource, it is possible to request explanations and clarifications to address knowledge gaps or initiate a knowledge review quiz covering the recommended resource. The former currently relies on the LLM assistant’s internal world knowledge but in future could be based on functions that provide well-motivated explanations, for example, using analogies that are identified as appropriate using learners’ known background knowledge \cite{ilievski99capturing,10.1162/tacl_a_00688}. Also, the behaviour here can extend from explanations to summarisation of previously consumed learning materials.

If a quiz is requested or agreed to, the assistant calls a function which generates a number of questions and returns them along with answers one at a time whereupon the LLM presents them to the user and scores the answers. Section \ref{sec:qg} explains how a micro-skill is developed to generate educational questions contextualised to specific Knowledge Components from a previously consumed learning resource. The ability to control the topical relevance of the questions allows the assistant to use the learner model to personalise the testing opportunity. In the initial version of the assistant, the results are simply captured for later use, but future versions should be able to use the results to update learners’ models appropriately. Also, the behaviour can be extended to the assistant using personalised quizzes to improve its confidence in learner skill estimates voluntarily, without the learner explicitly requesting a quiz.

\subsection{Engagement with Recommended Resources}
To recommend a resource, the assistant calls a function which recommends a resource from the content index by considering learners’ background knowledge, interests, their TrueLearn engagement model (see Section~\ref{sec:learner_model}), and immediate engagement history. This introduces the title of the video and sets the current recommendation link to view the video as well as making it possible for the LLM to call functions that retrieve additional information for the resource when relevant. This currently includes access to the raw transcript and KCs covered (from Wikification).  The learner thus has a number of options for engagement, including simply viewing the video or asking the assistant for a summary or key points covered after which they may, for example, choose to request a different recommendation without viewing the content. If the learner chooses to view the video, an engagement event is registered by the interface similar in convention to the events described in the PEEK Dataset~\cite{bulathwela2021peek} which is used to update a TrueLearn engagement model for the learner.

\subsection{Learner Model}
\label{sec:learner_model}

As per figure \ref{fig:logical_model}, \emph{interests} and \emph{experience} are the two main factors considered by the TrueReason system concerning user state. Resources are recommended based on learners’ interests and background knowledge. 
We use the TrueLearn model \cite{bulathwela2022sus}, one of our prior works, capable of capturing both learner interests/goals and knowledge as the learner model. Another key advantage of TrueLearn is its ability to capture learner knowledge/interest state in a humanly intuitive format \cite{truelear_python}, which is beneficial in building a human-centric AI system. A limitation of TrueLearn is the inability to recommend materials for a new user. In the proposed TrueReason system, the conversation interface initially converses with the learner to identity their current knowledge, interests and goals. The recommendations in the inception of the learner journey are based on resources that overlap with the set of topics of interest. At the same time, engagement signals are collected with the recommended learning resources to estimate the TrueLearn model – in the “cold start recommendation regime”. 


Once enough engagement events have been observed the TrueLearn model is used to make recommendations by means of calculating the engagement probability of resources in the content index with the TrueLearn novelty classifier – the “engagement model recommendation regime”. This method is similar to the \emph{switching-based hybrid recommendation} we have used with TrueLearn in prior work \cite{bulathwela2022sus}. In the prior work, we used a popularity-based prior \cite{bulathwela_edm_population} rather than a prior extracted by an LLM using conversation. This approach is designed to be a baseline approach which begins to address the cold start problem for new learners while a more advanced recommendation is developed. For more information about the development of a more advanced recommendation model that can serve as a drop-in replacement for this baseline, readers can refer to the the reinforcement learning educational recommender described in section \ref{sec:Reinforcement_Learning_ER}.

\section{Micro Skills} \label{sec:microskills}


\subsection{Reinforcement Learning Multi-step Educational Recommender}
\label{sec:Reinforcement_Learning_ER}
In Sect.~\ref{sec:learner_model} the TrueLearn novelty classifier is recommended as the baseline approach to address the cold start problem. This approach serves to establish an estimation of the knowledge components which form the learner's interest as well as their knowledge state. Once the learner's interests and knowledge state is established it is possible to create a trajectory of resources to recommend to the learner in order to improve their knowledge state over their interests. This is a non-trivial task given that the trajectory over the resources from one knowledge state to a desired knowledge state cannot always be achieved by simply recommending resources that directly involve the learner's interests. Many knowledge components in the the learners identified interests may require prerequisite knowledge components before a learner would be willing to engage with a recommended educational resource. This implies that there is a delayed reward for engaging with resources that don't necessarily directly involve the interest of the learner, but rather indirectly aid the learner to further engage with their interests. This creates a sequential decision-making problem as to which resource should be recommended. Having an expert to identify the prerequisites of any knowledge component will not always be possible. It is thus required that these trajectories be learned. However, it is seldom the case that we have enough data from learners to learn these trajectories, and thus we need a strategy to learn these trajectories. In this section a method for a learning such a viable trajectory through available educational resources is established using reinforcement learning.

\subsubsection{Markov Decision Process}
A general RL environment is setup as a Markov Decision Process (MDP), which consists of an agent interacting with an environment. The agent will receive an observation $o_{t}=O(s_{t})$ which is derived from the true state of the environment $s_{t}\in S$ at each time step $t$. The agent will perform an action, $a_{t}$ obtained from its policy $\pi(s_{t})$. The probability of the transition of the $s_{t}$ to $s_{t'}$ and $o_{t}$ to $o_{t'}$ is given by $T(s_{t'},o_{t'}|s_{t},a_{t})$. A reward $r_{t}$ is given at each time step and is given by the environments reward function $R(s_{t},a_{t})$. The goal of the agent is to maximise the total expected reward, $E[R_{t}] = \sum_{t'=t}^{\infty}\gamma^{t'-t}r_{t'}$, where $\gamma$ is the discount factor.

\subsubsection{Training Environment}
\label{sec:Training_Env}

In order to adapt the problem of learning a trajectory through a set of given resources we will need to establish an environment in which the recommender system (the agent) can learn which are appropriate resources to recommend. An approach of simply learning trajectories from known successful learner's activity could be used, however, data of that nature is not freely available in all cases and thus the need for a suitable proxy to generate such data arises. TrueLearn has the capability to predict engagement of a learner given enough interactions to estimate that learner's knowledge state. Thus, under the assumption that enough interactions have occurred to estimate the correct learner knowledge state, TrueLearn can be used proxy for learner engagement in the absence of learner engagement data. This gives TrueLearn the dual purpose of both estimating the learner's knowledge state and interests as well as the ability to be used as the proxy for learner engagement.

\begin{figure}[]
    \centering

    \includegraphics[width=\textwidth]{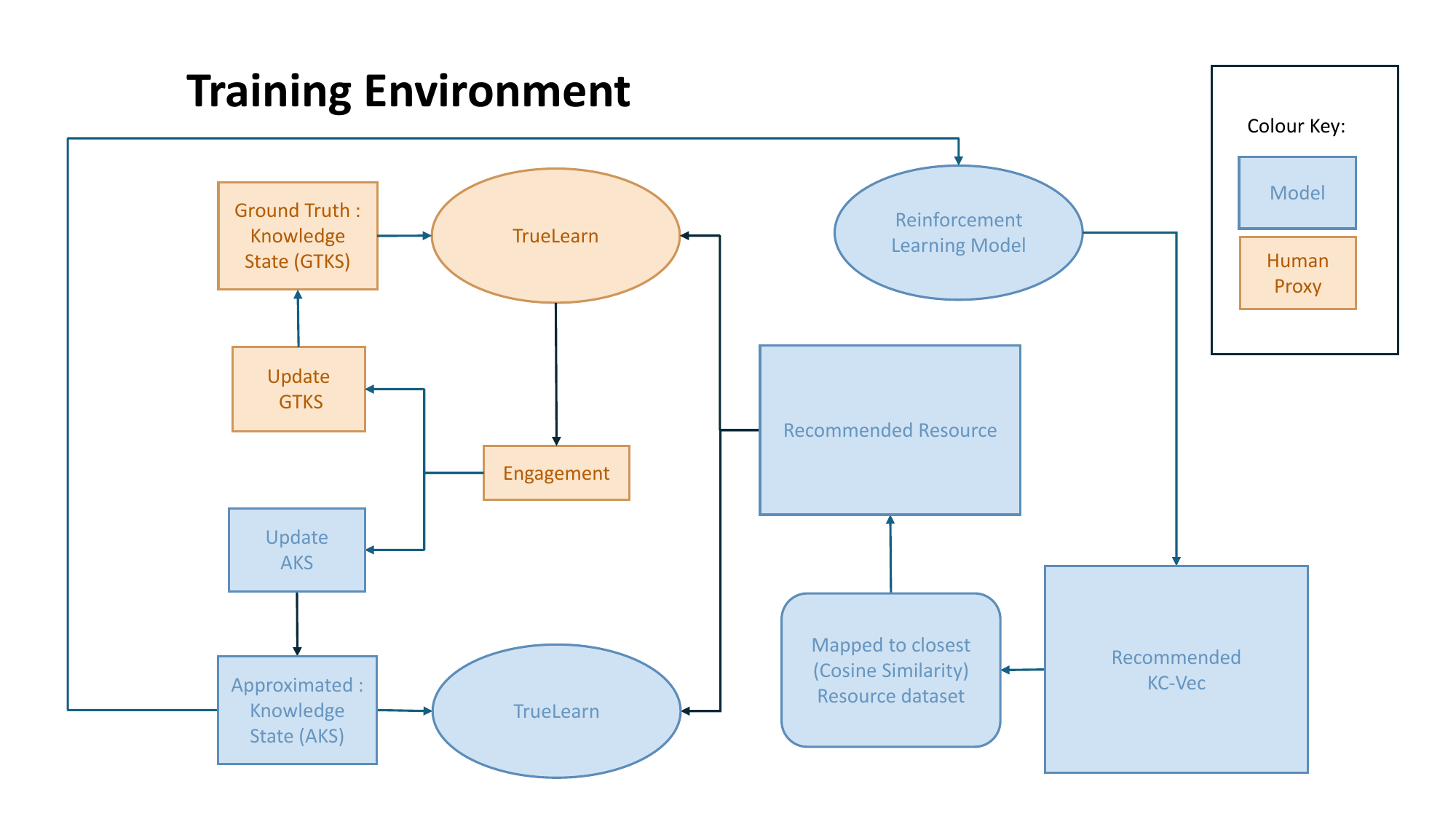}
    \caption{
    The architecture of the training environment using two instances of TrueLearn to act as the human proxy to simulate engagement and update the ground truth knowledge state (GTKS) in the first instance, and to update approximated knowledge state (AKS) in the second instance after being given a recommended educational resource from the RL algorithm. 
    }
    \label{fig:training_environment}
    \end{figure}

Taking the assumption that the ground truth knowledge state (GTKS) is known for a given learner then an instance of TrueLearn can be instantiated to act as an artificial learner and give a suitable indication of engagement as well as give a simulated update to the GTKS based on engagement of the resource. This allows the construction of an environment capable of simulating any number of artificial learners with known knowledge states and interests and, within the capabilities of TrueLearn, emulate engagement for a recommended educational resource. A second instance of TrueLearn can be used to obtain the approximated knowledge state (AKS) of the learner. Given such an environment it is then possible to use a suitable RL method to learn a policy which recommends a knowledge component vector (KC-Vec) which can then be mapped to educational resources, using Cosine Similarity, which a user would engage with. The capability to simulate the entire process of obtaining an AKS, recommending a resource, emulating engagement, and performing the respective updates to the GTKS and AKS based on engagement allows a suitably chosen RL method to learn entire trajectories over a set of available educational resources, with the aim of working toward an improved knowledge state. A general framework of this training environment can be seen in Fig \ref{fig:training_environment}.

This RL training environment can be setup as an MDP. The agent, the RL model, will receive the AKS as the observation $o_{t}=O(s_{t})$. The true state of the environment $s_{t}\in S$ will be the GTKS and interests of the learner. The chosen RL model will give the policy $\pi(s_{t})$ which will give a KC-Vec which will then be mapped to a resource in our available educational resource set, the knowledge components of the educational resource will then become the action, $a_{t}$. The transition of states in the environment from $s_{t}$ to $s_{t'}$ and $o_{t}$ to $o_{t'}$ will be given the updates of the GTKS and AKS by TrueLearn at each time step $t$. The reward function $R(s_{t},a_{t})$ can be set to the following in order to take the learners interests into account:
\begin{equation} \label{eq:RLRS_reward_function}
    R(s_{t},a_{t}) = \sum_{i\in I}(KC_{t}^{i}-KC_{t'}^{i}),
\end{equation}
where $KC_{t}^{i}$ and $KC_{t'}^{i}$ are the knowledge component values of the GTKS for the learner at time steps $t$ and $t'$ respectively for each interest $i\in I$ of the learner. This formulation makes it possible to work toward an increase in the knowledge state of each of the learners' interests.

\subsubsection{Deep Deterministic Policy Gradient}

\begin{figure}[]
    \centering

    \includegraphics[width=\textwidth]{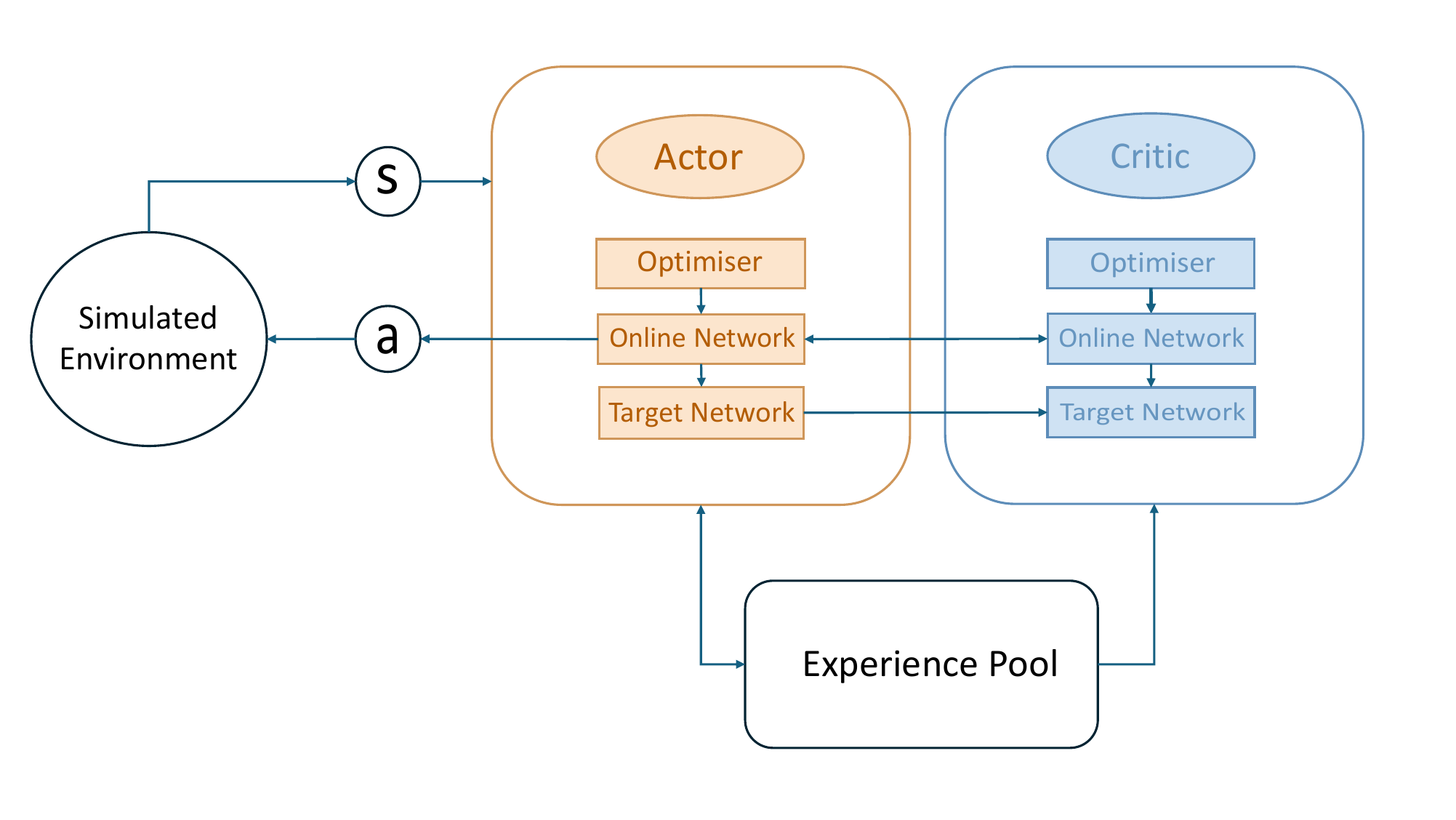}
    \caption{
    The Deep Deterministic Policy Gradient architecture, including the actor, the critic, the experience replay pool and the simulated environment.
    }
    \label{fig:DDPG}
    \end{figure}

Given the continuous nature of the state and actions in this environment an appropriate RL algorithm is that of Deep Deterministic Policy Gradient (DDPG) \cite{lillicrap2015continuous}. This is a off-policy algorithm, making it perfect for training from both simulated data and data collected from actual learners. This actor-critic model based off of the Deep Q-Leaning Networks (DQN). That is, it uses the concept from the Bellman equation:
\begin{equation}
    Q^{*}(s_{},a_{t})=E[r(s_{t},a_{t})+\gamma \max_{a_t'}Q^{*}(s_{t'},a_{t'})].
\end{equation}
As seen in Figure \ref{fig:DDPG}, this is a model designed to have an actor which interacts with the simulated training environment described in Sect. \ref{sec:Training_Env}, using its policy $\pi(o_t)$ to provide an action $a_{t}$, note, the policy uses the observation derived from the state, as the true state of the environment is not available to the actor. These these actions are then mapped to available learning resources, with the KC-vec related to the learning resource being stored as the action in the experience replay buffer (the experience pool). The critic, using the action created by the policy of the actor, as well as the replays from the experience replay buffer, is then used to obtain the value function $Q(s_t,a_t)$. The target networks in both the actor and the critic are used to aid in the stabilisation of learning the value function as well as help approximate future expected values of the value function based on the state at the next time step. These are updated to match the current online versions after a fixed number of time steps.

Using this framework we set up an actor with three linear fully connected layers, each followed be a ReLU function, with the last layer using a $\tanh$ to produce the recommended KC-vec. The critic uses three linear fully connected layers, each followed by a ReLU function, and the output of the last layer is used as the value function.

\subsubsection{Results}

\begin{figure}[]
    \centering

    \includegraphics[width=\textwidth]{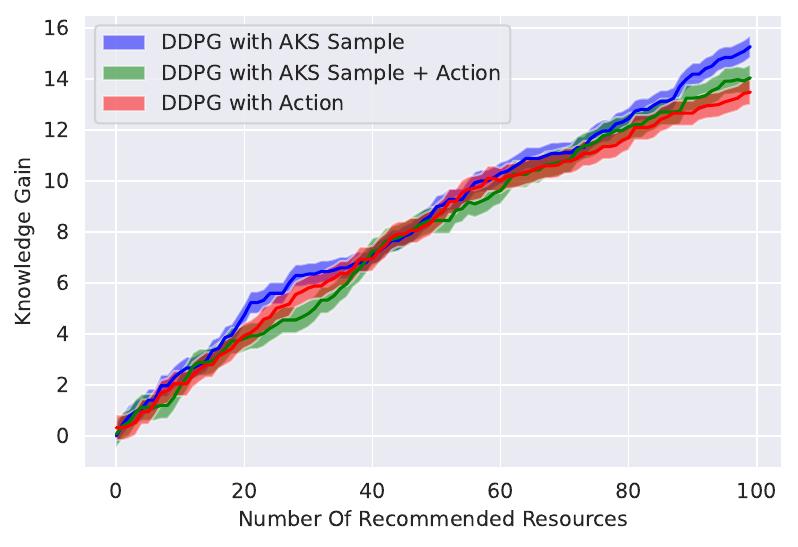}
    \caption{
    The Deep Deterministic Policy Gradient method applied to the training environment, including the three different models for the observations states. The results show the cumulative increase in in knowledge for a set of topics of interest after 50 000 training steps.
    }
    \label{fig:DDPG_Model_results}
    \end{figure}

To test the RL educational recommender, three experiments were run with the training environment. For each experiment we initialised a user with the same GTKS and the same set of interest topics, these were run 5 times for each model. The GTKS is used strictly for engagement prediction using the TrueLearn model and the reward function is calculated based on the change in the GTKS after updating the GTKS as per TrueLearns rules at each time step. 
Three variations of the DDPG model were then tested. The first model uses an input of KC-vecs sampled from from the AKS of the the proxy human with an assumed GTKS. This KC-vec is sampled at each time step and the AKS is updated by TrueLearn based on the predicted engagement from the GTKS instance of TrueLearn. The second model uses the concatenation of the KC-vec sampled from the AKS as well as the DDPG algorithm's suggested action KC-vec from the previous time step as inputs, for the initial time step, we simply use the KC-vec sampled from the AKS as the action KC-Vec. The third model uses only the action KC-vec at each time step, with the initial time step action simply being the KC-vec sampled from the AKS. As seen in Figure \ref{fig:DDPG_Model_results}, the DDPG model which only uses the AKS sample marginally outperforms the other two models.

\subsubsection{RL Recommender Conclusion}

The RL Recommender informed by TrueLearn as a proxy human has been shown to be capable of recommending a sequence of resources to increase a learners knowledge state in specific interests based on the engagement and updates done by the TrueLearn model. Future work entails checking the qualitative flow of the recommended resources and end ensure that there is a logical flow of topics, and not just arbitrary jumps between the learners interest topics.

\subsection{Automatic Topic-Controlled Educational Question Generation} \label{sec:qg}

Creating lesson materials and generating topic-specific, relevant, and age-appropriate questions for teaching have long been identified as time-intensive tasks for teachers, and an area where increased consistency is also expected to improve educational outcomes for students \cite{giannakos2024promise}. Although learning analytics and AI in Education researchers have long explored ways to support teachers' question generation capabilities through data-driven insights and models, attempts on \textbf{Topic-Controlled Question Generation (T-CQG)} have been less successful, primarily due to the lack of quality in the generated content. The use of language models, however, has the potential to address these quality concerns by leveraging recent advancements in natural language processing (NLP) for automatic educational question generation (EdQG). EdQG (and T-CQG) models can be integrated into personalised learning systems, to advance the system's capability to perform precise diagnostics on learner's knowledge gaps (providing contextualised formative/summative assessment). From the learners' points of view, prior research also suggests a strong correlation between the personalisation of testing and knowledge retention \cite{Bahrick1993}, which further supports the importance of topic-controlled question generation. T-CQG can serve as a small yet important micro skill that enhances TrueReason.

Educational Question Generation (QG) involves automatically generating questions from a specific text passage or a document. The main goal of QG is to produce questions that are not only syntactically and semantically correct but also contextually relevant and meaningful for the intended use. There has been a growing use of computational models to generate contextually relevant and grammatically correct questions \cite{wang2024llmseducation}. Early implementations of QG practices with data-driven approaches predominantly utilized sequence-to-sequence (seq2seq) architectures incorporating Recurrent Neural Networks (RNNs) \cite{Du2017}. More recently, the focus shifted towards employing end-to-end techniques facilitated by deep neural networks \cite{zhang2021review}. Topic-controlled question generation (T-CQG) is a specialised task in educational question generation where questions are generated constrained on a topic.  For instance, \cite{dathathri2020plug} and \cite{khalifa2021distributional} utilised GPT-2 combined with either an attribute classifier or training another autoregressive language model to guide the generated text towards a topic. However, these approaches typically generated content that is too broadly categorised (such as a category being 'science'), failing to achieve the level of topic specificity required for them to be of real-world value for educational practitioners. More contemporary models leverage pre-trained transformers like GPT \cite{blobsteinangel,elkins2024teachers}, however, these models use memory (instead of context) to generate questions, making it challenging to prescribe questions grounded on a very specific learning material (that the learner has seen in the past). Our prior work has shown that fine-tuning small Language models for educational question generation can perform as equally well as LLMs while improving scalability and sustainability \cite{fawzi2023small}. We hypothesise that a small language model can be finetuned for T-CQG via fine tuning. However, a scarcity of datasets for this task is a major challenge. The method we use to develop the T-CQG micro-skill entails creating a novel training dataset from existing public datasets using Wikification \cite{Brank2017}.

\subsubsection{Problem Definition}

Topic-controlled question generation takes a target topic in addition to the descriptive text as context into account while generating the models' outputs. 
To define topic-controlled question generation (T-CQG) precisely, let us suppose a learner $\ell$ has consumed study resources that containing knowledge $c$ linked to various topics $T_c$. TrueReason's goal is to generate questions $\hat{q}_t$, where $\hat{q}_t$ is an educational question about the target topic $t$, where $t \in T_c$, and $\hat{q}_t$ consists of a sequence of tokens $q_t \in \{w_1, \dots, w_{|q_t|}\}$ of arbitrary length $|q_t|$. The generated question not only requires to be contextually relevant to ${c}$, but align with the topic $t$. This task can be represented to identify the optimal question $\hat{q_t}$ that maximises the conditional probability as per equation \ref{eq:problem}.

\begin{equation} \label{eq:problem}
\hat{q_t} = \arg \max_{q_t} p({q_t} | {c},{t}) = \arg \max_{q_t} \sum_{i=1}^{|{q_t}|} \log p(w_{i} | {c}, {t}, w_1 \dots w_{i-1})    
\end{equation}

where, $p({q_t} | {c},{t})$ denotes the conditional probability that also depends on the tokens $w \in q_t$.

\subsubsection{Research Questions}\label{methodology} 

This work revolves around one main research question:
\begin{itemize}
    \item Is it feasible to fine-tune a pre-trained language model (PLM) to perform T-CQG?
\end{itemize}

\subsubsection{Datasets Utilised}\label{datasets_utilised}


We used the SQuAD 1.1, the Stanford Question Answering Dataset, comprising over 100,000 questions crafted by crowd workers based on a selection of 536 Wikipedia articles \cite{rajpurkar2016squad} as the source for creating new datasets (SQuAD+ and MixSQuAD as described in section \ref{sec:noveldata} below) for fine-tuning the models.

For evaluation, we used the KhanQ dataset \cite{gong-etal-2022-khanq} as it presents a more relevant challenge for educational question generation. It includes 1,034 high-quality questions in the STEM fields generated by learners, which aim to probe deep understanding of subjects taught in Khan Academy's online courses \footnote{\url{https://www.khanacademy.org}}. Despite its smaller size relative to SQuAD, KhanQ aligns more closely with our objective to generate topic-based and relevant educational questions (as per prior work \cite{fawzi2024humanlike}). To adapt the dataset for topic-based evaluation, we use the same approach as MixSQuAD (section \ref{sec:mixsquad}) to create a dataset with contrasting topic-based questions. We refer to the transformed version of the KhanQ dataset as \emph{MixKhanQ} dataset. 



\subsubsection{Creating Novel Datasets for T-CQG}\label{sec:noveldata} 

\begin{figure}[h]
\begin{center}
\includegraphics[width=.75\linewidth]{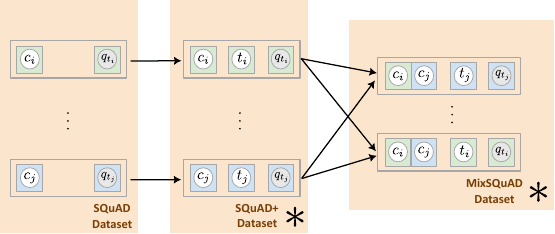}
\caption{Methodology for generating the different training datasets proposed in the model from the contexts $c$, topics $t$ and target questions $q_t$ (shaded as label) from the SQuAD dataset is illustrated using two random examples from the dataset, example $i$ (green) and example $j$ (blue). The contexts $c_i$ and $c_j$ glued together are concatenated texts treated as a single field in the dataset. The orange rectangles indicate the scope of the datasets while (*) marks the newly proposed datasets.}
\label{fig:datasets}
\end{center}
\end{figure}

A core contribution of this work is to introduce a novel data enrichment method that leads to the creation of new datasets that are derived from conventional question generation datasets. As described in \ref{datasets_utilised}, we derive the new datasets from SQuAD and KhanQ. These datasets already contain the context $c$ and the \emph{label} question $q_t$  from a human (contrast to $\hat{q}_t$ in equation \ref{eq:problem} which denotes the \emph{generated} question). We append an additional field to the dataset, \emph{Topic} $t$, and create three novel datasets, 1) SQuAD+ and 2) MixSQuAD for the T-CQG task. The process of generating the three datasets is presented in figure \ref{fig:datasets}.

\paragraph{\textbf{Linking the Target Topic to Data Points, SQuAD+ Dataset}}

To identify semantic annotations for every context and question, we employ wikification \cite{Brank2017}, which annotates text inputs with relevant concepts from Wikipedia ($T_c$). We retain the top 5 concepts for each text (context and question) based on their PageRank scores, which reflect the authority of the concept over the annotated text. To make sure that we can link the topical alignment between the question and the context, we only retain examples where at least one common Wikipedia concept is present between the context and the question pair (i.e. |$T_c \cap T_q| \geq 1$). We select the concept with the highest PageRank score in the question (most authoritative) as the target topic $t$. This method ensures that the most closely related annotation is selected as the topic for each pair, and confirms that the topic is appropriately aligned with both the context and the question, thus avoiding situations where the topic may be relevant to one but not the other. As a result, both datasets have been enhanced to include paragraph-level contexts $c$, identified topics $t$, and corresponding questions $q_t$, as shown in Fig.~\ref{fig:datasets}. 

\paragraph{\textbf{Introducing Pair-wise Contrastive Examples: MixSQuAD Dataset}}

We also create an enhanced dataset to synthesise a contrastive learning setting while fine-tuning the PLM for T-CQG leading to the \emph{MixSQuAD} dataset. When creating this dataset, we randomly pick pairs of observations from the SQuAD+ dataset. For each pair of examples $i$ and $j$ containing $(c_i, t_i, q_{t_i})$ and $(c_j, t_j, q_{t_j})$ respectively, we create two new examples where they share a common context $c_ic_j$ where the two contexts are concatenated. The data representation of the MixSQuAD dataset is presented in figure \ref{fig:datasets}. This approach aims to enhance the model's understanding of topics and the relationship between context, topic, and question by serving novel contrastive examples. An added benefit of the novel MixSQuAD dataset is that the context presented to the model during fine-tuning is guaranteed not to be previously encountered in the large corpora used for training foundational models. This method results in a diverse collection of 10,000 mixed data entries in the MixSQuAD dataset.

\subsubsection{Models} \label{sec:models}

With the relevant datasets created, we built two models to answer the research question. The models used in experiments are created by finetuning the \texttt{T5-Small} \cite{raffel2020exploring} model, a small Language Model (sLM) used in our prior work for educational question generation \cite{bulathwela2023scalable,fawzi2024humanlike}. We fine-tuned the foundational model
using the Adam optimizer with a batch size of 64, the learning rate of $1e-3$, and epsilon of $1e-8$. We use a maximum sequence length of 512 for the encoder, and 128 for the decoder. We train all models for a maximum of 50 epochs with an early stopping based on the validation loss. We conducted fine-tuning for T-CQG using the same finetuning approach used by \cite{martin2020} for controlling complexity in simplifying texts although our dataset creation method is very different.

\textbf{Baseline Model: } We used the proposed SQuAD+ dataset (described in section \ref{sec:noveldata}) to finetune the T5 PLM. 

\textbf{TopicQG Model: } The key difference between the baseline model and the proposed TopicQG model lies in the data used for fine-tuning the T5-small model. We introduced the TopicQG model to contrastive examples using the novel dataset created, MixSQuAD (described in section \ref{sec:noveldata}). Such mixed contexts, which may feature sentences with vastly differing concepts, are designed to enhance the T5 model's understanding of the semantic relationships between context $c$, topic $t$, and question $q_t$.

\subsubsection{Evaluation}

When evaluating the final models, we focused on the extent to which the generated question $\hat{q_t}$ is \emph{semantically related} to the prescribed topic $t$. We quantify this using semantic relatedness.
For measuring the semantic relatedness, $SemRel(q_t, \hat{q}_t)$, we needed metrics that can quantify the relatedness between the reference question $q_t$ and the generated question $\hat{q}_t$. We used the \texttt{BERTScore} \cite{zhang2020bertscore} and the Wikipedia-based Topic Semantic Relatedness (\texttt{WikiSemRel}) \cite{Tagme} metrics for these evaluations. 

\textbf{BERTScore} leverages BERT contextual embeddings of tokens to calculate the similarity between two text extracts, improving upon the traditional exact match methods. Our early experiments showed that the BERTSCore tend to inflate the similarity between $q_t$ and $\hat{q}_t$, as there are words like "what" and "why" that overlap even if the generated question is not about the salient topic $t$ of the reference question. Therefore, we excluded stopwords in the reference and generated questions prior to calculating the BERTScore. BERTSCore is a score in the range (0,1) where 0 indicates no relatedness.

\textbf{WikiSemRel} quantifies the semantic relatedness between the Wikipedia-based concepts extracted from the reference question $q_t$ and the generated question $\hat{q}_t$. We employ the WAT API \cite{WAT2014} service to calculate semantic relatedness using the Jaccard-based measure, that uses the outward links to other Wikipedia pages to calculate similarity \cite{Ponza2020}. We Wikify the generated question to compute the WikiSemRel score which is within range (0,1) where 0 indicates no relatedness. 

\paragraph{\textbf{Experimental Setup for Automated Performance Evaluations}}

\begin{figure}[h]
\begin{center}
\centerline{\includegraphics[width=.7\linewidth]{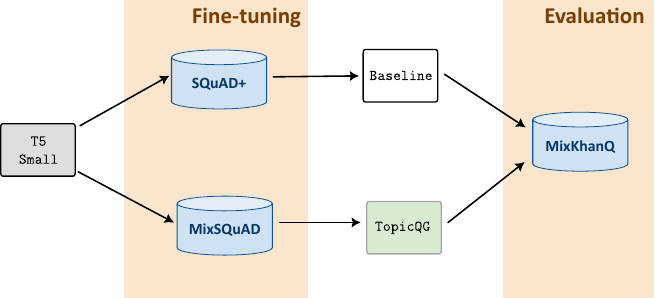}}
\caption{Methodology for training and evaluating the Baseline model (black) and TopicQG model (green). The shaded grey box indicates that the \texttt{T5-Small} model was available pre-trained prior to the experiments while the non-shaded models contained parameters trained during the experiments.}
\label{fig:experiment}
\end{center}

\end{figure}

Figure \ref{fig:experiment} illustrates the experimental setup designed to address RQs 2-5. 
A total of six models (including TopicQG's base, 8bit, and 4bit versions) have been developed as described in detail in section \ref{sec:models} and represented as coloured boxes in figure \ref{fig:experiment}. Each model is evaluated using the MixKhanQ dataset.

\subsection{Results and Discussion}

In this section, we present the results from the experiments described in section \ref{methodology}. 
Table \ref{tab:topic} presents the semantic closeness between the prescribed topic and the generated questions. Table \ref{tab:examples} presents a random set of topic-controlled question generations based on the context text provided in five different subject areas (Computing, Economics, Chemistry, Art, and Biology).

\begin{table}[]
\caption{Semantic relatedness between the generated questions $\hat{q}$ on (i) prescribed topic ${t}$ vs. (i) alternative topic ${t'}$ and the reference question on the prescribed topic ${q}_{t}$. The best performance for each metric is highlighted in \textbf{bold}.}
\label{tab:topic}
\begin{tabular}{p{2cm}|p{1.25cm}p{1.25cm}p{2cm} p{1.25cm}p{1.25cm}p{2cm}} \hline
    Model                & \multicolumn{3}{c}{BERTScore} & \multicolumn{3}{c}{WikiSimRel} \\
                         &  $\hat{q}_t \uparrow$ & $\hat{q}_{t'} \downarrow$ & $|\hat{q}_t - \hat{q}_{t'}| \uparrow$ &$ \hat{q}_t \uparrow$ & $\hat{q}_{t'} \downarrow$ & $|\hat{q}_t - \hat{q}_{t'}| \uparrow$ \\ \hline
Baseline                 & \textbf{0.859}   & {0.859}            & {0.000}         & {0.615}          & {\textbf{0.070}} & 0.545          \\
TopicQG                  & \textbf{0.859}   & \textbf{0.830}     & \textbf{0.029}         & \textbf{0.727}   & {0.132}         & \textbf{0.595} \\\hline
\end{tabular}
\end{table}

\begin{table}[]
\begin{center}
\caption{Randomly Selected Sample Generations from the TopicQG Model}
\centerline{\includegraphics[width=1.1\linewidth]{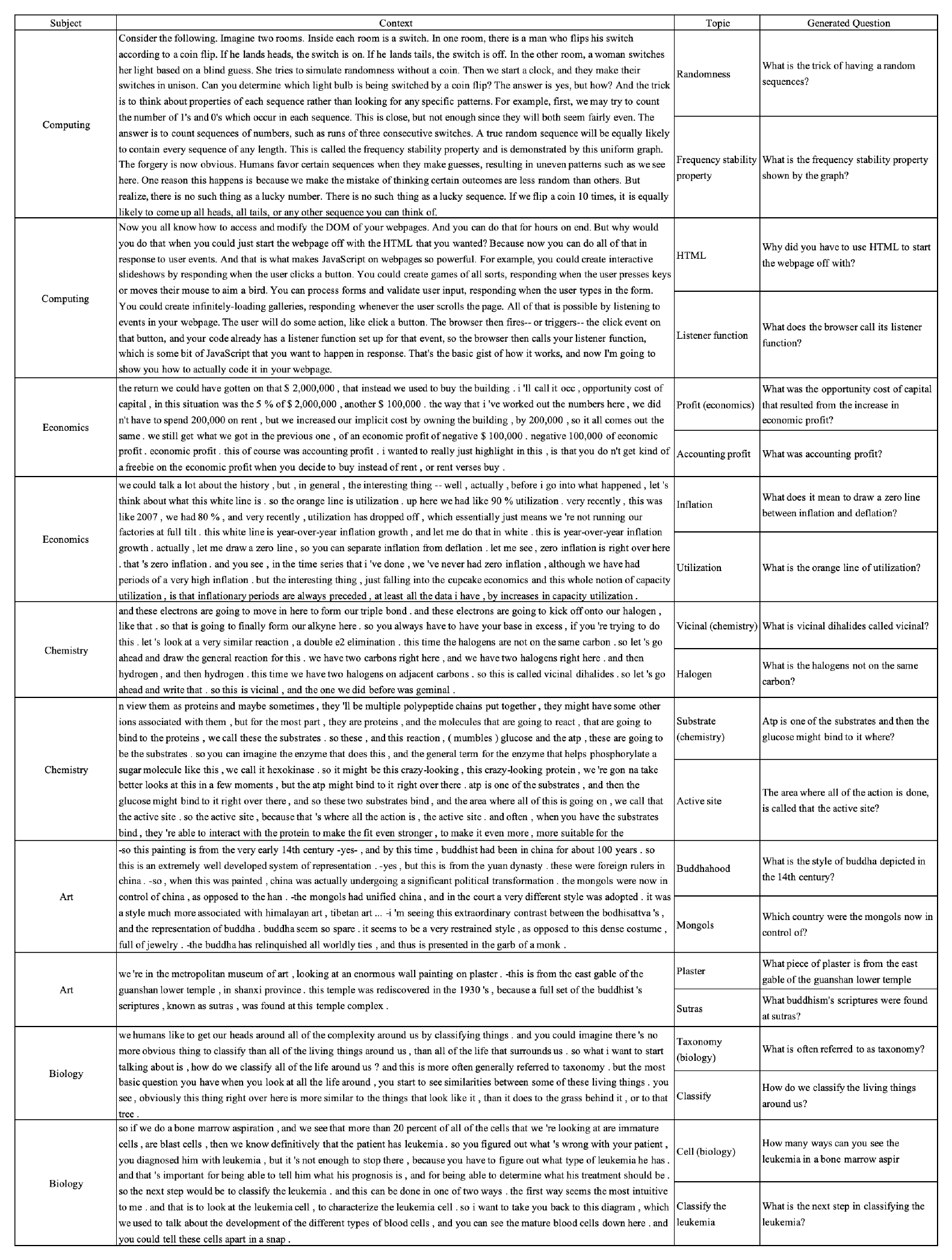}}
\label{tab:examples}
\end{center}
\end{table} 


Table \ref{tab:topic} provides us an indication of the degree to which the generated question $\hat{q}_t$ resembles the reference question $q_t$. This is a proxy for topical relevance as the reference question is implicitly aligned with the controlled topic. The perplexity of generations does not raise significant concerns over the quality of generations as the random examples in table \ref{tab:examples} doesn't indicate visible signs of deterioration. However, it is evident that the questions generated in table \ref{tab:examples} lack qualities like answerability, meaning further improvements are needed before plugging the generations automatically in a personalised learning system.

\subsubsection{Topic-Controlled EdQG Conclusion}

This paper proposes a novel approach to fine-tuning pre-trained language models to effectively address the challenge of generating topic-controllable questions based on paragraph-level context within educational settings. The model proposed here has the potential to decrease teacher workload and improve personalised learning platforms. It can also be scaled at a minimal cost in a safe and ethical manner to decrease the over-reliance of academic researchers on commercial LLMs like ChatGPT. 

\section{Implications of the Results for Research and Practice} \label{sec:dis}

The work outlined in this chapter demonstrates how a group of independent AI models can be composed together to articulate advanced cognitive behaviour that entails higher-order reasoning and planning. 

\subsection{Society-first AI}

A powerful element of the proposed system is that the intelligence demonstrated by the system is caused by a multitude of AI models than a monolithic large language model. Such an architecture can bring different benefits in terms of both sustainability and functional excellence.  

\begin{itemize}
    \item The complexity of maintaining the individual small models is significantly small, improving accessibility, control and ownership capabilities for both models and data. 
    \item The generalisation capabilities of individual micro-skills can be thoroughly tested and their generalisation capabilities can be better guaranteed
    \item The risk of unlearning previous skills with further training on new skills can be eliminated due to the separation of micro-skills
    \item  The modular architecture allows “turning off” certain skills contextually (e.g. models can be \emph{left behind} when entering low resources episodes, temporary privacy/security zones etc.)
    \item Different AI systems (data-driven vs. rule-based, blackbox vs.transparent etc.) can be composed together in one system.
    \item The micro-skills can be composed in different combinations to orchestrate different generation behaviours without additional training
\end{itemize}

\subsection{Cross-Modal, Cross-Lingual, Cross-Device Lifelong Learning}

The ability to modularise the micro skills packaged with TrueReason allows expanding the repertoire of skills that TrueReason can incorporate over time. The ability to use an extensive collection of learning materials and learning behaviours that span across multiple languages and modalities in the future positions the proposal to support lifelong learning across a diverse learner population with diverse goals. The use of Wikipedia as the universal knowledge representation (content model) supports this vision due to its cross-linguality and ability to evolve at scale \cite{su16020781}. While a conversational interface has been implemented for TrueReason currently, the multimodal learning activities that TrueReason can potentially support  in the future will extend its user interfaces to more sophisticated learning interfaces (E.g. AI-powered discovery \cite{x5learn,perez2022watch}.

\subsection{Towards Sentient Learning Companions}

The notion of \emph{society of minds} is adapted in the TrueReason learning assistant to create an educational assistant that can compose different skills to synthesise an activity-rich, engaging learning journey. In the context of education, such an architecture for a lifelong learning agent is quite useful as different learning episodes of a persons life can demand different learning experiences. As lifelong learning entails transitioning through different subject domains over a long period of time, a domain model and a collection of learning resources that can support such coverage should be accessible to the system \cite{su16020781}. With the flexibility of micro-skills, the requirements to have monolythic domain models, learner models and document corpora that support all possible learning scenarios are no longer required as subsets of specialised learner models, domain models and content collections can be utilised by the reasoning agent depending on the context (e.g. subject domain). Same goes to the expressivity of the individual micro-skills. While transparent learning models and decision components in intelligent tutors can build trust and trigger meta-cognitive activity among self-regulated learners \cite{williamson2021effects}, such transparency is not required in all parts of the system. Specifically, neural generative AI components can provide much richer and user friendly outputs than their transparent counterparts \cite{deep_kt}. Due to this reason, the micro-skill approach strikes great balance between the desired functional and non-functional requirements of a humane-AI system. 

With TrueReason architecture, the AI system is able to incorporate a battery of micro-skills. Tools that can understand graph/chart images and code already exist. With the right micro-skills, TrueReason will understand and appropriately incorporate visual, auditory and various other forms of data modalities in learning journeys. Our previous work has shown how a TrueLearn-like model can be used to infer aesthetic preferences of learners (e.g. readability level, complexity of vocabulary, length of content etc.) and select personalised learning materials that also satisfies non-topical preferences of learners \cite{molan2020accessibility}. The T-CQG work described in section \ref{sec:qg} shows how micro-skills can be created to generate content driven by specific constrains (topics/concepts in this case). We hypothesise that a series of micro-skills can be built i) for generating different forms of content (e.g. summaries, explanations, distractors for questions etc.) with ii) additional constraints (e.g. readability level, difficulty, length and even aspects like background noise level of the learning environment). As an example, TrueReason can go beyond exclusively topic-driven personalisation and consider the affective state of the learner to calibrate the tone of language and readability level. It can consider your preference of politics over sports when presenting analogies. It can take into account the background noise level of the learner and decide to use a text activity than a video. 

Also, the superset of micro-skills available to the AI system can expand over time as the planning and reasoning LLM can learn to use new micro-skills introduced to the system. This would allow TrueReason to evolve its behaviour over time. One could argue that the TrueReason's micro-skills allows it to see, hear and feel the learner, adapting its behavior to accommodate these sensory cues while evolving its bahavior over time, embodying sentient aspects to the learning system.   

\section{Conclusion} \label{sec:conclusion}

In this chapter, we presented TrueReason, an exemplar personalised learning system that is developed by ochestrating a collection of AI models together to act as a society of minds. Contrary to developing a monolythic large langauge model that is continously trained to acquire new skills over time, the proposed modular structure allows new skill aquisition to be less computationally demanding with no risk of unlearning/forgetting. The generalisation guarantees of each micro skill is also improved while the architecture also allows "plug-and-play" capability. The TrueReason system developed at present leverages Open AI's ChatGPT with Assistant API to provide a conversation like interface to the learner while several micro skills relating to prerequisite/next steps identification, open learner modelling, question generation and multi-step recommendation are already implemented. The system shows the ability to orchestrate the diverse micro skills using the ChatGPT LLM that operationalises reasoning, planning and continuing the conversation with the human learner. The system opens up many possibilities to the future in terms of creating more micro skills that focus on different pedagogical interventions while opportunities to train the planning agent to support advance pedagogical interventions.      

\subsection{Future Work}

Looking into the future, the immediate focus can be on how additional micro skills leveraging generative AI can be developed for summarisation, explanation/feedback generation and presenting analogies that are contextualised for topical content/ complexity of vocabulary and various other state-based attributes of individual learners. Simultaneously, it should be explored how the planning/reasoning agents can be made smaller (to make them computationally efficient) and more sophisticated (using specialised datasets such as debating transcripts).

\bibliographystyle{splncs04}
\bibliography{chapter}

\begin{thebibliography}{10}
\providecommand{\url}[1]{\texttt{#1}}
\providecommand{\urlprefix}{URL }
\providecommand{\doi}[1]{https://doi.org/#1}

\bibitem{abdelrahman2023knowledge}
Abdelrahman, G., Wang, Q., Nunes, B.: Knowledge tracing: A survey. ACM
  Computing Surveys  \textbf{55}(11),  1--37 (2023)

\bibitem{bauman2018recommending}
Bauman, K., Tuzhilin, A.: Recommending remedial learning materials to students
  by filling their knowledge gaps. MIS Quarterly  \textbf{42}(1),  313--A7
  (2018)

\bibitem{blei2003latent}
Blei, D.M., Ng, A.Y., Jordan, M.I.: Latent dirichlet allocation. Journal of
  Machine Learning Research  \textbf{3} (2003)

\bibitem{blobsteinangel}
Blobstein, A., Izmaylov, D., Yifat, T., Levy, M., Segal, A.: Angel: A new
  generation tool for learning material based questions and answers. In:
  Proc.~of the NeurIPS Workshop on Generative AI for Education (GAIED)

\bibitem{Brank2017}
Brank, J., Leban, G., Grobelnik, M.: Annotating documents with relevant
  wikipedia concepts. In: Proc. of Slovenian KDD Conference on Data Mining and
  Data Warehouses (SiKDD) (2017)

\bibitem{truelearn}
Bulathwela, S., Perez-Ortiz, M., Yilmaz, E., Shawe-Taylor, J.: Truelearn: A
  family of bayesian algorithms to match lifelong learners to open educational
  resources. In: AAAI Conference on Artificial Intelligence (2020)

\bibitem{x5learn}
Bulathwela, S., Kreitmayer, S., P\'{e}rez-Ortiz, M.: {What's in It for Me?
  Augmenting Recommended Learning Resources with Navigable Annotations}. In:
  Proc.~of the Int. Conf. on Intelligent User Interfaces Companion (2020)

\bibitem{bulathwela2023scalable}
Bulathwela, S., Muse, H., Yilmaz, E.: Scalable educational question generation
  with pre-trained language models. In: Proc.~of Int. Conf. on Artificial
  Intelligence in Education. pp. 327--339. Springer (2023)

\bibitem{bulathwela2021peek}
Bulathwela, S., P{\'{e}}rez{-}Ortiz, M., Novak, E., Yilmaz, E., Shawe{-}Taylor,
  J.: {PEEK:} {A} large dataset of learner engagement with educational videos.
  CoRR  \textbf{abs/2109.03154} (2021), \url{https://arxiv.org/abs/2109.03154}

\bibitem{bulathwela2021truelearn}
Bulathwela, S., P{\'{e}}rez{-}Ortiz, M., Yilmaz, E., Shawe{-}Taylor, J.:
  {Semantic TrueLearn: Using Semantic Knowledge Graphs in Recommendation
  Systems}. CoRR  \textbf{abs/2112.04368} (2021),
  \url{https://arxiv.org/abs/2112.04368}

\bibitem{semantic_truelearn}
Bulathwela, S., P{\'e}rez-Ortiz, M., Yilmaz, E., Shawe-Taylor, J.: Leveraging
  semantic knowledge graphs in educational recommenders to address the
  cold-start problem. In: Semantic AI in Knowledge Graphs, pp. 1--20. CRC Press
  (2023)

\bibitem{su16020781}
Bulathwela, S., Pérez-Ortiz, M., Holloway, C., Cukurova, M., Shawe-Taylor, J.:
  Artificial intelligence alone will not democratise education: On educational
  inequality, techno-solutionism and inclusive tools. Sustainability
  \textbf{16}(2) (2024)

\bibitem{bulathwela_edm_population}
Bulathwela, S., Verma, M., Ortiz, M.P., Yilmaz, E., Shawe-Taylor, J.: Can
  population-based engagement improve personalisation? {A} novel dataset and
  experiments. In: Mitrovic, A., Bosch, N. (eds.) Proceedings of the 15th
  International Conference on Educational Data Mining. pp. 414--421.
  International Educational Data Mining Society (July 2022).
  \doi{10.5281/zenodo.6853185}

\bibitem{bulathwela2022sus}
{Bulathwela, Sahan and P{\'e}rez-Ortiz, Mar{\'\i}a and Yilmaz, Emine and
  Shawe-Taylor, John}: {Power to the Learner: Towards Human-Intuitive and
  Integrative Recommendations with Open Educational Resources}. Sustainability
  \textbf{14}(18) (2022)

\bibitem{bull2008metacognition}
Bull, S., Kay, J.: Metacognition and open learner models. In: The 3rd workshop
  on meta-cognition and self-regulated learning in educational technologies, at
  ITS2008. pp. 7--20 (2008)

\bibitem{Bull2016}
Bull, S., Kay, J.: Smili: a framework for interfaces to learning data in open
  learner models, learning analytics and related fields. International Journal
  of Artificial Intelligence in Education  \textbf{26}(1),  293--331 (2016)

\bibitem{cachola-etal-2020-tldr}
Cachola, I., Lo, K., Cohan, A., Weld, D.: {TLDR}: Extreme summarization of
  scientific documents. In: Cohn, T., He, Y., Liu, Y. (eds.) Findings of the
  Association for Computational Linguistics: EMNLP 2020. pp. 4766--4777.
  Association for Computational Linguistics, Online (Nov 2020).
  \doi{10.18653/v1/2020.findings-emnlp.428},
  \url{https://aclanthology.org/2020.findings-emnlp.428}

\bibitem{nn_know_comp}
Chaplot, D.S., MacLellan, C., Salakhutdinov, R., Koedinger, K.: Learning
  cognitive models using neural networks. In: Proc.~of Artificial Intelligence
  in Education (2018)

\bibitem{Corbett1994}
Corbett, A.T., Anderson, J.R.: Knowledge tracing: Modeling the acquisition of
  procedural knowledge. User Modeling and User-Adapted Interaction
  \textbf{4}(4) (1994)

\bibitem{dathathri2020plug}
Dathathri, S., Madotto, A., Lan, J., Hung, J., Frank, E., Molino, P., Yosinski,
  J., Liu, R.: Plug and play language models: A simple approach to controlled
  text generation. In: International Conference on Learning Representations
  (2020), \url{https://openreview.net/forum?id=H1edEyBKDS}

\bibitem{denny2024generative}
Denny, P., Gulwani, S., Heffernan, N.T., K{\"a}ser, T., Moore, S., Rafferty,
  A.N., Singla, A.: Generative ai for education (gaied): Advances,
  opportunities, and challenges. arXiv preprint arXiv:2402.01580  (2024)

\bibitem{Du2017}
Du, X., Shao, J., Cardie, C.: Learning to ask: Neural question generation for
  reading comprehension. In: Proc. Annual Meeting of the Association for
  Computational Linguistics. pp. 1342--1352 (2017)

\bibitem{elkins2024teachers}
Elkins, S., Kochmar, E., Cheung, J.C., Serban, I.: How teachers can use large
  language models and bloom's taxonomy to create educational quizzes. In: AAAI
  Conference on Artificial Intelligence (2024)

\bibitem{fawzi2024humanlike}
Fawzi, F., Balan, S., Cukurova, M., Yilmaz, E., Bulathwela, S.: Towards
  human-like educational question generation with small language models. In:
  Artificial Intelligence in Education. Posters and Late Breaking Results,
  Workshops and Tutorials, Industry and Innovation Tracks, Practitioners,
  Doctoral Consortium and Blue Sky. Communications in Computer and Information
  Science, vol.~2150. Springer, Cham (2024)

\bibitem{fawzi2023small}
Fawzi, F., Amini, S., Bulathwela, S.: Small generative language models for
  educational question generation. In: Proc.~of the NeurIPS Workshop on GAIED

\bibitem{Tagme}
Ferragina, P., Scaiella, U.: Tagme: on-the-fly annotation of short text
  fragments (by wikipedia entities). In: Proceedings of the 19th ACM
  International Conference on Information and Knowledge Management. p.
  1625–1628. CIKM '10, Association for Computing Machinery, New York, NY, USA
  (2010). \doi{10.1145/1871437.1871689},
  \url{https://doi.org/10.1145/1871437.1871689}

\bibitem{gong-etal-2022-khanq}
Gong, H., Pan, Liangming~andHu, H.: {KHANQ}: A dataset for generating deep
  questions in education. In: Proceedings of the 29th International Conference
  on Computational Linguistics (2022)

\bibitem{hooshyar2020open}
Hooshyar, D., Pedaste, M., Saks, K., Leijen, {\"A}., Bardone, E., Wang, M.:
  Open learner models in supporting self-regulated learning in higher
  education: A systematic literature review. Computers \& Education
  \textbf{154},  103878 (2020)

\bibitem{ilievski99capturing}
Ilievski, F., Shenoy, K., Klein, N., Chalupsky, H., Szekely, P.: Capturing
  concept similarity with knowledge graphs. Manuscrit de Th{\`e}se page
  \textbf{99}

\bibitem{ilkou2021}
Ilkou, E., Abu-Rasheed, H., Tavakoli, M., Hakimov, S., Kismih{\'o}k, G., Auer,
  S., Nejdl, W.: Educor: An educational and career-oriented recommendation
  ontology. In: The Semantic Web -- ISWC 2021. pp. 546--562. Springer
  International Publishing, Cham (2021)

\bibitem{jia2024llm}
Jia, Q., Cui, J., Du, H., Rashid, P., Xi, R., Li, R., Gehringer, E.:
  Llm-generated feedback in real classes and beyond: Perspectives from students
  and instructors. In: Proceedings of the 17th International Conference on
  Educational Data Mining. pp. 862--867 (2024)

\bibitem{KANG201752}
Kang, J., Lee, H.: Modeling user interest in social media using news media and
  wikipedia. Information Systems  \textbf{65} (2017).
  \doi{https://doi.org/10.1016/j.is.2016.11.003}

\bibitem{khalifa2021distributional}
Khalifa, M., Elsahar, H., Dymetman, M.: A distributional approach to controlled
  text generation. In: International Conference on Learning Representations
  (2021), \url{https://openreview.net/forum?id=jWkw45-9AbL}

\bibitem{klavsnja2018social}
Kla{\v{s}}nja-Mili{\'c}evi{\'c}, A., Ivanovi{\'c}, M., Vesin, B., Budimac, Z.:
  Enhancing e-learning systems with personalized recommendation based on
  collaborative tagging techniques. Applied Intelligence  \textbf{48},
  1519--1535 (2018)

\bibitem{li2023evaluating}
Li, X., Henriksson, A., Duneld, M., Nouri, J., Wu, Y.: Evaluating embeddings
  from pre-trained language models and knowledge graphs for educational content
  recommendation. Future Internet  \textbf{16}(1), ~12 (2023)

\bibitem{lillicrap2015continuous}
Lillicrap, T.: Continuous control with deep reinforcement learning. arXiv
  preprint arXiv:1509.02971  (2015)

\bibitem{Lindsey2014}
Lindsey, R.V., Khajah, M., Mozer, M.C.: Automatic discovery of cognitive skills
  to improve the prediction of student learning. In: Ghahramani, Z., Welling,
  M., Cortes, C., Lawrence, N.D., Weinberger, K.Q. (eds.) Advances in Neural
  Information Processing Systems 27, pp. 1386--1394. Curran Associates, Inc.
  (2014)

\bibitem{martin2020}
Martin, L., Villemonte~de La~Clergerie, {\'E}., Sagot, B., Bordes, A.:
  {Controllable Sentence Simplification}. In: {LREC 2020 - 12th Language
  Resources and Evaluation Conference}. Marseille, France (May 2020),
  \url{https://inria.hal.science/hal-02678214}

\bibitem{miladi2024comparative}
Miladi, F., Psych{\'e}, V., Lemire, D.: Comparative performance of gpt-4,
  rag-augmented gpt-4, and students in moocs. In: International Conference on
  Breaking Barriers with Generative Intelligence. pp. 81--92. Springer (2024)

\bibitem{modran2024llm}
Modran, H., Bogdan, I.C., Ursuțiu, D., Samoila, C., Modran, P.L.: Llm
  intelligent agent tutoring in higher education courses using a rag approach
  (2024)

\bibitem{molan2020accessibility}
Molan, M., Bulathwela, S., Orlic, D.: Accessibility recommendation system. In:
  Proceedings of the OER20: Open Education Conference (2020)

\bibitem{moutsinas2021graph}
Moutsinas, G., Shuaib, C., Guo, W., Jarvis, S.: Graph hierarchy: a novel
  framework to analyse hierarchical structures in complex networks. Scientific
  Reports  \textbf{11}(1),  13943 (2021)

\bibitem{park2024empowering}
Park, M., Kim, S., Lee, S., Kwon, S., Kim, K.: Empowering personalized learning
  through a conversation-based tutoring system with student modeling. In:
  Extended Abstracts of the CHI Conference on Human Factors in Computing
  Systems. pp. 1--10 (2024)

\bibitem{perez2022watch}
P{\'e}rez~Ortiz, M., Bulathwela, S., Dormann, C., Verma, M., Kreitmayer, S.,
  Noss, R., Shawe-Taylor, J., Rogers, Y., Yilmaz, E.: Watch less and uncover
  more: Could navigation tools help users search and explore videos? In:
  Proceedings of the 2022 Conference on Human Information Interaction and
  Retrieval. pp. 90--101 (2022)

\bibitem{piaopredicting}
Piao, G.: Recommending knowledge concepts on mooc platforms with
  meta-path-based representation learning. In: Proc.~of Int. Conf. on
  Educational Data Mining (2021)

\bibitem{piao_social_networks}
Piao, G., Breslin, J.G.: Analyzing aggregated semantics-enabled user modeling
  on google+ and twitter for personalized link recommendations. In: Proceedings
  of the 2016 Conference on User Modeling Adaptation and Personalization. UMAP
  '16 (2016)

\bibitem{piao_mooc}
Piao, G., Breslin, J.G.: Analyzing mooc entries of professionals on linkedin
  for user modeling and personalized mooc recommendations. In: Proceedings of
  the 2016 Conference on User Modeling Adaptation and Personalization. UMAP '16
  (2016)

\bibitem{wat_api}
Piccinno, F., Ferragina, P.: From tagme to wat: A new entity annotator. In:
  Proc.~of the First Int. Workshop on Entity Recognition \& Disambiguation. ERD
  '14 (2014)

\bibitem{WAT2014}
Piccinno, F., Ferragina, P.: From tagme to wat: a new entity annotator. In:
  Proceedings of the First International Workshop on Entity Recognition \&
  Disambiguation. p. 55–62. ERD '14, Association for Computing Machinery
  (2014). \doi{10.1145/2633211.2634350},
  \url{https://doi.org/10.1145/2633211.2634350}

\bibitem{deep_kt}
Piech, C., Bassen, J., Huang, J., Ganguli, S., Sahami, M., Guibas, L.J.,
  Sohl-Dickstein, J.: Deep knowledge tracing. In: Advances in Neural
  Information Processing Systems (2015)

\bibitem{piro2024mylearningtalk}
Piro, L., Bianchi, T., Alessandrelli, L., Chizzola, A., Casiraghi, D.,
  Sancassani, S., Gatti, N.: Mylearningtalk: An llm-based intelligent tutoring
  system. In: International Conference on Web Engineering. pp. 428--431.
  Springer (2024)

\bibitem{ponza_semantic_relate}
Ponza, M., Ferragina, P., Chakrabarti, S.: On computing entity relatedness in
  wikipedia, with applications. Knowledge-Based Systems  \textbf{188} (2020)

\bibitem{Ponza2020}
Ponza, M., Ferragina, P., Chakrabarti, S.: On computing entity relatedness in
  wikipedia, with applications. Knowledge-Based Systems  \textbf{188} (2020)

\bibitem{truelear_python}
Qiu, Y., Djemili, K., Elezi, D., Shalman~Srazali, A., Pérez-Ortiz, M., Yilmaz,
  E., Shawe-Taylor, J., Bulathwela, S.: A toolbox for modelling engagement with
  educational videos. Proceedings of the AAAI Conference on Artificial
  Intelligence  \textbf{38}(21),  23128--23136 (2024).
  \doi{10.1609/aaai.v38i21.30358},
  \url{https://ojs.aaai.org/index.php/AAAI/article/view/30358}

\bibitem{raffel2020exploring}
Raffel, C., Shazeer, N., Roberts, A., Lee, K., Narang, S., Matena, M., Zhou,
  Y., Li, W., Liu, P.J., et~al.: Exploring the limits of transfer learning with
  a unified text-to-text transformer. J. Mach. Learn. Res.  \textbf{21}(140),
  1--67 (2020)

\bibitem{rajpurkar2016squad}
Rajpurkar, P., Zhang, J., Lopyrev, K., Liang, P.: Squad: 100,000+ questions for
  machine comprehension of text (2016), \url{https://arxiv.org/abs/1606.05250}

\bibitem{reicherts2022}
Reicherts, L., Park, G.W., Rogers, Y.: Extending chatbots to probe users:
  Enhancing complex decision-making through probing conversations. In:
  Proceedings of the 4th Conference on Conversational User Interfaces. CUI '22,
  Association for Computing Machinery, New York, NY, USA (2022).
  \doi{10.1145/3543829.3543832}

\bibitem{sampson2011ask}
Sampson, D.G., Zervas, P., Kalamatianos, A.: Ask-lost 2.0: A web-based tool for
  social tagging digital educational resources in learning environments. Social
  Media Tools and Platforms in Learning Environments pp. 387--398 (2011)

\bibitem{schick2024toolformer}
Schick, T., Dwivedi-Yu, J., Dess{\`\i}, R., Raileanu, R., Lomeli, M., Hambro,
  E., Zettlemoyer, L., Cancedda, N., Scialom, T.: Toolformer: Language models
  can teach themselves to use tools. Advances in Neural Information Processing
  Systems  \textbf{36} (2024)

\bibitem{shi2024chain}
Shi, Z., Gao, S., Chen, X., Feng, Y., Yan, L., Shi, H., Yin, D., Chen, Z.,
  Verberne, S., Ren, Z.: Chain of tools: Large language model is an automatic
  multi-tool learner. arXiv preprint arXiv:2405.16533  (2024)

\bibitem{shi2024learning}
Shi, Z., Gao, S., Chen, X., Feng, Y., Yan, L., Shi, H., Yin, D., Ren, P.,
  Verberne, S., Ren, Z.: Learning to use tools via cooperative and interactive
  agents. arXiv preprint arXiv:2403.03031  (2024)

\bibitem{shin2021saint+}
Shin, D., Shim, Y., Yu, H., Lee, S., Kim, B., Choi, Y.: Saint+: Integrating
  temporal features for ednet correctness prediction. In: LAK21: 11th
  International Learning Analytics and Knowledge Conference. pp. 490--496
  (2021)

\bibitem{10.1162/tacl_a_00688}
Sourati, Z., Ilievski, F., Sommerauer, P., Jiang, Y.: {ARN: Analogical
  Reasoning on Narratives}. Transactions of the Association for Computational
  Linguistics  \textbf{12},  1063--1086 (09 2024)

\bibitem{stamper2024enhancing}
Stamper, J., Xiao, R., Hou, X.: Enhancing llm-based feedback: Insights from
  intelligent tutoring systems and the learning sciences. In: International
  Conference on Artificial Intelligence in Education. pp. 32--43. Springer
  (2024)

\bibitem{SyedC17}
Syed, R., Collins{-}Thompson, K.: Optimizing search results for human learning
  goals. Inf. Retr. J.  \textbf{20}(5),  506--523 (2017)

\bibitem{thus2024exploring}
Th{\"u}s, D., Malone, S., Br{\"u}nken, R.: Exploring generative ai in higher
  education: a rag system to enhance student engagement with scientific
  literature. Frontiers in Psychology  \textbf{15},  1474892 (2024)

\bibitem{wang2024llmseducation}
Wang, S., Xu, T., Li, H., Zhang, C., Liang, J., Tang, J., Yu, P.S., Wen, Q.:
  Large language models for education: A survey and outlook (2024),
  \url{https://arxiv.org/abs/2403.18105}

\bibitem{williamson2021effects}
Williamson, K., Kizilcec, R.F.: Effects of algorithmic transparency in bayesian
  knowledge tracing on trust and perceived accuracy. International Educational
  Data Mining Society  (2021)

\bibitem{xiong2016going}
Xiong, X., Zhao, S., Van~Inwegen, E.G., Beck, J.E.: Going deeper with deep
  knowledge tracing. International Educational Data Mining Society  (2016)

\bibitem{Yudelson13}
Yudelson, M.V., Koedinger, K.R., Gordon, G.J.: Individualized bayesian
  knowledge tracing models. In: Lane, H.C., Yacef, K., Mostow, J., Pavlik, P.
  (eds.) Proc.~of Artificial Intelligence in Education (2013)

\bibitem{zarrinkalam2020extracting}
Zarrinkalam, F., Faralli, S., Piao, G., Bagheri, E.: Extracting, mining and
  predicting users’ interests from social media  (2020)

\bibitem{zhang2021review}
Zhang, R., Guo, J., Chen, L., Fan, Y., Cheng, X.: A review on question
  generation from natural language text. Trans. on Information Systems
  \textbf{40}(1),  1--43 (2021)

\bibitem{zhang2020bertscore}
Zhang, T., Kishore, V., Wu, F., Weinberger, K.Q., Artzi, Y.: Bertscore:
  Evaluating text generation with bert (2020),
  \url{https://arxiv.org/abs/1904.09675}

\end{thebibliography}
\end{document}